\documentclass[12pt,letterpaper]{article}%
\usepackage{amsfonts}
\usepackage{amsmath}
\usepackage{amssymb}
\usepackage{graphicx}%
\providecommand{\U}[1]{\protect\rule{.1in}{.1in}}

\begin{document}
\begin{center}
{\bf\LARGE Observations on BI from $\mathcal{N}=2$ Supergravity\\ and the General Ward Identity} \\
\vskip 2 cm
{\bf \large Laura Andrianopoli$^{1,2}$, Patrick Concha$^{1,2,3}$, Riccardo D'Auria$^{1,2}$,\\
Evelyn Rodriguez$^{1,2,3}$ and Mario Trigiante$^{1,2}$}
\vskip 8mm
 \end{center}
\noindent {\small{$^1$ DISAT, Politecnico di Torino, Corso Duca
    degli Abruzzi 24, I-10129 Turin\\
    $^{2}$  \it Istituto Nazionale di
    Fisica Nucleare (INFN) Sezione di Torino, Italy\\
    $^{3}$ Departamento de F\'{\i}sica, Universidad de
Concepci\'{o}n, Casilla 160-C, Concepci\'{o}n, Chile}

\vskip 1.5 cm
\begin{center}
{\small {\bf Abstract}}
\end{center}
The multi-vector generalization of a rigid, partially-broken $\mathcal{N}=2$ supersymmetric theory is presented as a rigid limit of a suitable gauged $\mathcal{N}=2$ supergravity with electric, magnetic charges and antisymmetric tensor fields. This on the one hand generalizes a known result by Ferrara, Girardello and Porrati while on the other hand allows to recover the multi-vector BI models of \cite{Ferrara:2014oka} from $\mathcal{N}=2$ supergravity as the end-point of  a hierarchical limit in which the Planck mass first and then the supersymmetry breaking scale are sent to infinity.
 We define, in the parent supergravity model, a new symplectic frame in which, in the rigid limit, manifest symplectic invariance is preserved and the electric and magnetic Fayet-Iliopoulos terms are fully originated from the dyonic components of the embedding tensor.
The supergravity origin of several features of the resulting rigid supersymmetric theory are then elucidated, such as the presence of a traceless ${\rm SU}(2)$- Lie algebra term in the Ward identity and the existence of a central charge in the supersymmetry algebra which manifests itself as a harmless gauge transformation on the gauge vectors of the rigid theory; we show that this effect can be interpreted as a kind of ``superspace non-locality'' which does not affect the rigid theory on space-time.
 To set the stage of our analysis we take the opportunity in this paper to provide and prove the
  relevant identities of the most general dyonic gauging of Special-Kaehler and Quaternionic-Kaehler isometries in a generic $\mathcal{N}=2$ model, which include the supersymmetry Ward identity, in a fully symplectic-covariant formalism.

\vskip 1 cm
\vfill
\noindent {\small{\it
    E-mail:  \\
{\tt laura.andrianopoli@polito.it}; \\
{\tt patrickconcha@udec.cl}; \\
{\tt riccardo.dauria@polito.it}; \\
{\tt everodriguez@udec.cl}; \\
{\tt mario.trigiante@polito.it}}}
   \eject
   \numberwithin{equation}{section}
   \section{Introduction}

Much attention has been recently devoted to the Born-Infeld (BI) theory and its multi-vector generalization, in relation to supersymmetric theories. These non-linear theories emerge from   a low-energy limit of partially-broken ${\rm U}(1)^n$  rigid  $\mathcal{N}=2$ supersymmetric theory \cite{Hughes:1986dn}, in which the supersymmetry breaking scale is sent to infinity \cite{Deser:1980ck,Cecotti:1986gb,Ferrara:2014oka}. This mechanism, as it was originally shown by \cite{APT} (APT model), requires the introduction of \emph{magnetic} Fayet-Iliopoulos (FI) terms  besides the electric ones, with the condition that the dual FI terms be not mutually local.  On the other hand the rigid, partially-broken  $\mathcal{N}=2$ theory with one vector multiplet of \cite{APT} was obtained as a rigid limit of a suitable $\mathcal{N}=2$ supergravity in \cite{FGP}. This defines a $\mathcal{N}=2$ supergravity origin of the one-vector BI theory.\par
The aim of our investigation is to embed the partially-broken, rigid  $\mathcal{N}=2$ theory of $n$ (abelian) vector multiplets in supergravity. This would elucidate the supergravity origin of the multifield BI theory of \cite{Ferrara:2014oka} and, in particular, to understand the origin of the dyonic FI as deriving from electric and magnetic charges in the supergravity gauged model. \par
In the original rigid limit devised in \cite{FGP}, the gauging was electric and partial supersymmetry breaking required the use of a specific choice of symplectic frame in which the prepotential of the special geometry does not exist. More general, partially-broken $\mathcal{N}=2$ supergravities were constructed in \cite{Fre:1996js} using an analogous choice of symplectic frame. This restriction, which is forced within the framework of standard (i.e. electric) gaugings by some no-go theorems \cite{Witten:1982df,Cecotti:1984rk}, can be avoided in the context of dyonic gaugings \cite{Dall'Agata:2003yr,D'Auria:2004yi,deWit:2005ub,de Wit:2007mt,Andrianopoli:2007ep}. It was indeed shown in \cite{Louis:2009xd} that partial supersymmetry breaking can be achieved in any symplectic frame (and in particular in  one in which the prepotential does exist) using an \emph{embedding tensor} \cite{Cordaro:1998tx,Nicolai:2000sc,deWit:2002vt}  with both electric and magnetic components. Consistency of such gaugings requires the introduction of antisymmetric tensor fields dual to scalars \cite{Dall'Agata:2003yr,D'Auria:2004yi,deWit:2005ub,de Wit:2007mt,Andrianopoli:2007ep}.\par

\par General electric-magnetic gaugings of $\mathcal{N}=2$ supergravity have been constructed in the  framework of superconformal calculus in \cite{deWit:2011gk}. A generic gauged $\mathcal{N}=2$ Poincar\'{e} supergravity can then be obtained from this analysis by suitably fixing the superconformal symmetry. However a direct construction of the most general electric-magnetic gaugings in $\mathcal{N}=2$ Poincar\'{e} supergravity, using a coordinate independent, manifestly symplectic-covariant description of the special-K\"ahler manifold, along the lines of \cite{Andrianopoli:1996cm}, is still missing.

The general form of the gauge-invariant bosonic lagrangian, using the embedding tensor formulation, was given  in \cite{deWit:2005ub} while specific abelian gaugings were constructed in \cite{Dall'Agata:2003yr,D'Auria:2004yi}.\footnote{in reference \cite{D'Auria:2004yi} also non-abelian gaugings were considered, however only of electric type.}
In this paper, to set the stage for the construction of the gauged model generalizing  that of \cite{FGP}, we make a step forward in this direction and give, in a self-contained form, all the relevant identities related to the most general gauging of special K\"ahler and quaternionic  K\"ahler isometries in a generic $\mathcal{N}=2$ model. Some of these identities are known, other were proven only for electric gaugings \cite{D'Auria:1990fj,Andrianopoli:1996cm} or within superconformal calculus \cite{deWit:2011gk}. Here we collect them and give for them a compact proof,
for generic dyonic gaugings, based on the coordinate-independent, symplectic-covariant description of the \emph{local} special-geometry and on the general constraints on the embedding tensor. Among these identities, a prominent role in our analysis will be played by the potential Ward-identity \cite{ward} \cite{D'Auria:2001kv}, which is required by the supersymmetry invariance of the gauged action. It follows from the quadratic constraints on the embedding tensor and a proof of it within $\mathcal{N}=2$ Poincar\'{e} supergravity, for a generic dyonic gauging, has been missing so far. Besides the definition of the rigid limit yielding a partially-broken $\mathcal{N}=2$  rigid supersymmetric theory of $n$ abelian vector multiplets, the general proof of the  Ward-identity for generic dyonic gaugings is a further result of our work. In order to present it in a self-contained fashion, we review in the Appendices the basic definitions and properties related to (local) special K\"ahler and quaternionic  K\"ahler manifolds.\par

The starting point of our analysis is then the construction of a suitable dyonic gauging of an $\mathcal{N}=2$ supergravity coupled to $n$ vector multiplets
and to hypermultiplets which allows for the definition of a rigid limit to a multi-vector APT model, thus generalizing \cite{FGP}.\\ The definition of a rigid limit in a $\mathcal{N}=2$ supergravity is not unique and is in general a subtle issue \cite{Andrianopoli:1996cm,Gunara:2013rca}: Rescalings of the fields and of the embedding tensor by powers of the ratio $\mu=M_{Pl}/\Lambda$ of the Planck mass $M_{Pl}$ to the supersymmetry breaking scale $\Lambda$ have to be devised in order for the original supersymmetries to survive the limit $\mu\rightarrow \infty$. Defining such a limit is an important part of our analysis.\par
The supergravity origin of the rigid theory is made manifest through  some characteristic results of the limiting procedure:
 First of all, although they decouple for $M_{Pl}\rightarrow \infty$, the gravitini and the hyperini (the fermion fields in the hypermultiplets) have a role in defining the general features of the resulting partially-broken rigid supersymmetry: Their supersymmetry transformation laws survive the rigid limit and contribute a non-trivial traceless constant matrix $C_A{}^B$ to the scalar potential Ward identity of the final supersymmetric theory:
\begin{equation}
\mathcal{V}\delta_A^B+C_A{}^B=\sum_{i=1}^n \delta \lambda^{iB}\delta \lambda_{iA}\,,\label{rigward0}
\end{equation}
where $\mathcal{V}$ is the scalar potential and $\lambda^{iA}$ and $\lambda_{iA}\equiv g_{i\bar{\jmath}}\,\lambda^{\bar{\jmath}}_A$ are the chiral and anti-chiral components of the gaugini. The constant matrix $C_A{}^B$, which in \cite{FGP} was put in relation to a central extension of the supersymmetry current algebra, is an essential ingredient in order for the partial supersymmetry breaking to occur in the rigid theory.  In \cite{FGP} it was shown, for a one-vector-multiplet model, that (\ref{rigward0}) originate from the supergravity Ward identity and that partial supersymmetry breaking in the rigid theory can occur even if supersymmetry is completely broken in  the \emph{hidden sector}, consisting of the decoupled gravitational multiplet and hypermultiplets. We show
the same feature in our generalized dyonic setting.

Moreover, a direct generalization of the  construction in \cite{FGP} to $n$ vector multiplets leads us to relate the FI terms of the rigid theory partly to (dyonic) components of the embedding tensor, and partly to constants entering the metric of the scalar manifold.
As we shall show, by an appropriate (electric) symplectic rotation we can  reformulate the theory in a symplectic frame where the supergravity interpretation of the FI terms is more transparent:
In this new frame, as opposed to the original one, in performing  the rigid limit  manifest symplectic invariance (which reduces from $Sp(2n+2)$ to $Sp(2n)$) is preserved and the electric and magnetic FI terms of the resulting theory fully originate from the components of the embedding tensor and not from constants entering the geometry of the scalar manifold. More specifically, if we denote by $A^\Lambda_\mu=(A^0_\mu,\,A^I_\mu)$, the $n+1$ supergravity vector fields, in the new symplectic frame, $A^0_\mu$ is consistently identified with the graviphoton while  $A^I_\mu$ with the vector fields of the resulting rigid theory. Moreover,  denoting by $\Theta_\Lambda{}^m$ the components of the embedding tensor which define the gauge generators ${\rm X}_\Lambda$ in terms of the isometry generators $t_m$ of the scalar manifold and by $\Theta^{\Lambda\,m}$ their magnetic counterparts, consistency of the supergravity gauging requires the following \emph{locality condition} to be satisfied \cite{Dall'Agata:2003yr,D'Auria:2004yi,deWit:2005ub}:
\begin{equation}\Theta^{\Lambda [m} \Theta_\Lambda{}^{n]}=0\,,\quad \Lambda =(0,I)=0,1,\cdots n\,.\label{locc}\end{equation}
In the rigid limit the electric and magnetic FI terms can be directly identified with $\Theta_I{}^m$ and $ \Theta^{I\,m}$, respectively, and the gauging is such that
\begin{equation}\label{nonloc}
   \Theta^{I [m} \Theta_I{}^{n]}=-\Theta^{0 [m} \Theta_0{}^{n]}\neq 0\,.
\end{equation}
The fact that in the supergravity framework $\Theta^{I [m} \Theta_I{}^{n]}$ fails to vanish, however, does not imply a failure of locality in the rigid theory on space-time. Indeed it turns out that on space-time the theory is perfectly local, the aforementioned  ``non-locality'' being confined to superspace, thereby  posing  no obstruction to a correct definition of the vector fields $A^I_\mu$ in the rigid theory
 which we shall discuss in Section 4. There we will explicitly show
 an interesting mechanism which is at work in the rigid limit. It is related to the well known property  of magnetic gaugings in supergravity  that the vector fields $A^\Lambda_\mu$ corresponding to non-vanishing magnetic components $\Theta^{\Lambda m}$ of the embedding tensor, are not well defined since the corresponding field strengths $F^\Lambda_{\mu\nu}$ are not covariantly closed \cite{Dall'Agata:2003yr,D'Auria:2004yi,deWit:2005ub,de Wit:2007mt,Andrianopoli:2007ep}
 \begin{equation}
 DF^\Lambda\propto \Theta^{\Lambda m}\,dB_m+\dots\neq 0\,,
 \end{equation}
$B_{m|\mu\nu}$ being antisymmetric tensor fields. This poses no problem because such vector fields, in a vacuum, are ``eaten'' by the tensor ones $B_m$ by virtue of the ``anti-Higgs'' mechanism \cite{Cecotti:1987qr}.  This is the case of the vectors  $A^I_\mu$ which are thus not well defined in the chosen supergravity gauging.
In the rigid limit however, as we shall show, the antisymmetric tensor fields decouple, thus preventing the anti-Higgs mechanism from taking place, so that the vectors  $A^I_\mu$ survive and, at the same time, become well defined.\\
As we shall illustrate in the same Section, the \emph{magnetic} character of the FI parameters $\Theta^{I\,m}$ in the rigid theory can be also related, besides to their position within the ${\rm Sp}(2n,\mathbb{R})$-covariant parameter vector $(\Theta_I{}^m,\,\Theta^{I\,m})$, to the following feature of the vector field-strengths: While $dF^I$ vanish in space-time, they \emph{do not} vanish in superspace since:
\begin{equation}
   d F^I =\frac{i}{2} \Theta^{I m} \mathcal P^x_m\,(\sigma^x)_A{}^B\,\bar{\psi}_B\wedge\gamma_a \psi^A\wedge V^a\neq 0\,.
   \end{equation}
This equation is the superspace counterpart of the fact that on space-time the commutator of two supersymmetries acts on the gauge field $A_\mu^I$ as a harmless gauge transformation, as stressed in reference \cite{Ferrara:1995gu}.

\par

The paper is organized as follows:\\
In Sect. 2 we give the general proof of the Ward identity for a generic electric-magnetic gauging of $\mathcal{N}=2$ supergravity. We also comment on its rigid limit for the specific gauging to be dealt with in the subsequent sections;\\
In Sect. 3 we give a generalization of the analysis in \cite{FGP} in order to  derive a partially-broken $\mathcal{N}=2$ rigid supersymmetric theory of $n$ abelian vector multiplets from a gauged $\mathcal{N}=2$ supergravity with electric and magnetic charges. We also derive the rigid Ward identity from the supergravity one;\\
In Sect. 4 we start from a different symplectic frame in which the supergravity origin of the electric and magnetic FI terms resulting from the rigid limit is more transparent. The issue of non-locality associated with the magnetic FI terms is also discussed;\\
In Sect. 5 the rigid limit is discussed in detail and performed on the supergravity Lagrangian, thus obtaining the multi-vector generalization of the APT model.\\
In Appendix A we review the definition and properties of (local) special K\"ahler and quaternionic  K\"ahler manifolds, giving for the latter a simple geometric characterization of the momentum maps associated with their isometries in the homogeneous symmetric case.\\
In Appendix B we prove some symplectic-covariant identities related to the general gauging of isometries of  local special K\"ahler manifolds. We also give the computational details of the proof of the Ward identity;\\
In Appendix C we summarize our rescaling prescription for the definition of the rigid limit.

\section{General $\mathcal{N}=2$ Gauging Identities} \label{gauge}
The aim of the present section is to give and prove identities which hold for the most general gauging of $\mathcal{N}=2$
supergravity involving both electric and magnetic charges. These include the Ward identity \cite{ward} which is required by the supersymmetry
invariance of the gauged Lagrangian. We shall derive these identities, as it was done in $\mathcal{N}>2$ models (see, for instance, \cite{Schon:2006kz,de Wit:2007mt}) from linear and quadratic constraints on the embedding tensor defining the gauge group.\par
The most general electric-magnetic gauging was considered in $\mathcal{N}=2$ conformal supergravity in \cite{deWit:2011gk}. Here we shall work in Poincar\'{e} supergravity using the symplectic covariant description of the special K\"ahler manifold and generalize the identities given in \cite{Andrianopoli:1996cm} to electric-magnetic gaugings and the analysis in \cite{Dall'Agata:2003yr} to non-abelian gauge groups. We believe it is useful to give, in this context, a comprehensive discussion of the identities which are relevant to the most general gauging, some of which are not present in the literature. These results will then be applied, in the later sections,  to the very specific electric-magnetic abelian gaugings in which the rigid limit of spontaneously broken  $\mathcal{N}=2$ supergravity is discussed. Some of the new relations presented here require rather technical proofs; the proofs will be explicitly given in Appendix \ref{ident}, leaving in the text only the corresponding results.
\par
We start from an $\mathcal{N}=2$ supergravity coupled to $n$ vector multiplets and $n_H$ hypermultiplets. The scalar sector consists of $n$ complex scalars $z^i$ and $4n_H$ hyperscalars $q^u$ parametrizing a special K\"ahler manifold $\mathcal{M}_{SK}$ \cite{Strominger:1990pd,D'Auria:1990fj,Ceresole:1995jg} and a quaternionic K\"ahler manifold $\mathcal{M}_{QK}$ \cite{Bagger:1983tt,Hitchin:1986ea,Galicki:1986ja}, respectively, so that the scalar manifold has the form:
\begin{equation}
\mathcal{M}_{scal}=\mathcal{M}_{SK}\times \mathcal{M}_{QK}\,.
\end{equation}
We refer the reader to \cite{Andrianopoli:1996cm} for a self-contained review of the properties of special K\"ahler and quaternionic K\"ahler manifolds. We recall the main concepts in Appendix \ref{SK}.\par
\paragraph{Some relevant relations of the sigma-model geometry.}
A special K\"ahler manifold is locally described by a choice of complex coordinates $z^i$ and a section of the flat holomorphic bundle defined on it:
\begin{equation}
\Omega^M(z)=\begin{pmatrix}X^\Lambda(z)\cr F_\Lambda(z)\end{pmatrix}\,\,,\,\,\,\Lambda=0,\dots, n\,,\quad M=1,\cdots ,2n+2\,,
\end{equation}
in terms of which the K\"ahler potential reads:
\begin{equation}
\mathcal{K}(z,\bar{z})=-\log[i\,\overline{\Omega}(\bar{z})^T\mathbb{C}\Omega(z)]\,,   \mbox{ where } \mathbb{C}
^{MN}=\left(
\begin{array}
[c]{cc}%
0 & 1\\
-1 & 0
\end{array}
\right)\,.\label{Kom}
\end{equation}
In terms of $\Omega$ and $\mathcal{K}$ one defines the covariantly holomorphic section $V^M\equiv e^{\frac{\mathcal{K}}{2}}\,\Omega^M$, see Appendix \ref{SK}, which transforms under a K\"ahler transformation (\ref{fOm}), (\ref{fK}) through a ${\rm U}(1)$ transformation (\ref{fV}).\par

A holomorphic function $f_g(z)$ and a constant symplectic matrix $\mathbb{M}[g]=(\mathbb{M}[g]_M{}^N)$ are associated with each element $g$ of the identity-connected component $G_{SK}$ of the  isometry group of  $\mathcal{M}_{SK}$  such that, if $g: z^i\rightarrow z'^i=z'^i(z)$:
\begin{equation}
\Omega(z')=e^{f_g(z)}\,\mathbb{M}[g]^{-T}\,\Omega(z)\,\,\Leftrightarrow\,\,\,\mathcal{K}(z',\bar{z}')=\mathcal{K}(z,\bar{z})-f_g(z)-\bar{f}_g(\bar{z})\,,\label{OmK}
\end{equation}
where $\mathbb{M}^{-T}\equiv (\mathbb{M}^{-1})^T$. If $\{t_a\}$ are the infinitesimal generators of $G_{SK}$ and $k_a=k_a^i(z) \partial_i+k_a^{\bar{\imath}}(\bar{z})\partial_{\bar{\imath}}$ the corresponding Killing vectors satisfying the closure conditions:
\begin{equation}
[t_a,\,t_b]=f_{ab}{}^c\,t_c\,\,,\,\,\,[k_a,\,k_b]=-f_{ab}{}^c\,k_c\,,
\end{equation}
equations (\ref{OmK}) imply:
\begin{eqnarray}
\label{Lie1}\ell_a\mathcal{K}&=&k_a^i \partial_i\mathcal{K}+k_a^{\bar{\imath}}\partial_{\bar{\imath}}\mathcal{K}=-(f_a+\bar{f}_a)\\
\label{Lie2}\ell_a\Omega^M&=&k_a^i\partial_i\Omega^M=-t_{aN}{}^M\,\Omega^N+f_a(z)\Omega^M\,,\\
\label{Lie3}\ell_a V^M&=&(k_a^i \partial_i+k_a^{\bar{\imath}}\partial_{\bar{\imath}})V^M=-t_{aN}{}^M\,V^N+\frac{f_a-\bar{f}_a}{2}\,V^M\,,\label{kV}\,,
\end{eqnarray}
where $f_a=\partial_if\,k^i_a$ and $t_{aN}{}^M$ is the symplectic matrix representation of the generator $t_a$ on covariant vectors: $t_{a[N}{}^P \mathbb{C}_{M]P}=0\,,\,\,(t_a\Omega)^M=-t_{aN}{}^M\,\Omega^N$.\par
Let us denote by $\mathcal{P}_a(z,\bar{z})$ the momentum map corresponding to $k_a$, defined as follows \cite{D'Auria:1990fj}:
\begin{equation}
k_a^i=i\,g^{i\bar{\jmath}}\,\partial_{\bar{\jmath}}\mathcal{P}_a\,,\,
\,\,k_a^{\bar{\imath}}=-i\,g^{\bar{\imath}i}\,\partial_{i}\mathcal{P}_a\,,\label{kpal}
\end{equation}
and satisfying, under general assumptions on $G_{SK}$ \cite{D'Auria:1990fj},
\begin{equation}
ig_{i\bar{\jmath}}\,k^i_{[a}\,k^{\bar{\jmath}}_{b]}=-\frac{1}{2}\,f_{ab}{}^c\,(\mathcal{P}_c-C_c)\,,\label{equivar1}
\end{equation}
where  $C_c$ is constant vector in the adjoint of $G_{SK}$ which can be reabsorbed by a  redefinition of $\mathcal{P}_c$. In what follows we shall make this redefinition: $\mathcal{P}_c-C_c\rightarrow \mathcal{P}_c$.\par
Eqs. (\ref{kpal}) are solved by (see Appendix A) :
\begin{eqnarray}\label{totti}
\mathcal{P}_a&=&-\frac{i}{2}\,\left(k_a^i \partial_i\mathcal{K}-k_a^{\bar{\imath}}\partial_{\bar{\imath}}\mathcal{K}\right)+{\rm Im}(f_a)=\nonumber\\
&=&i\,k_a^{\bar{\imath}}\partial_{\bar{\imath}}\mathcal{K}+ i\, \bar{f}_a=-i\,k_a^i \partial_i\mathcal{K}- i\, f_a\,,\label{Pa}
\end{eqnarray}
On the other hand, using (\ref{kV}) and (\ref{Pa}) we find:
\begin{equation}
k_a^i\,U_i^M=-t_{aN}{}^M\,V^N+i\,\mathcal{P}_a\,V^M\,.\label{kaUi}
\end{equation}
Contracting the above equation with $\mathbb{C}\overline{V}$ and using the special geometry relations $V^T\mathbb{C} \overline{V}=i,\,V^T\mathbb{C}U_i=0$, see Appendix \ref{SK}, we find:
\begin{equation}
\mathcal{P}_a=-V^N\,t_{aNM}\overline{V}^M=-\overline{V}^N\,t_{aNM}\,V^M\,,\label{PVtV}
\end{equation}
where we have defined $t_{aNM}\equiv t_{aN}{}^P\mathbb{C}_{PM}=t_{aMN}$. \par
Let us now prove the general property \cite{deWit:1984pk,deWit:2011gk}:
\begin{equation}
t_{a\,MN}\Omega^M\Omega^N=0\,\,,\,\,\,\,\forall t_a\,.\label{tOmOm}
\end{equation}
This property immediately follows by contracting  (\ref{kaUi}) with $\mathbb{C}\Omega$ and using the third of (\ref{defre}), i.e. $V^T\mathbb{C}U_i=0$, which implies
\begin{equation}
\Omega^T\mathbb{C} \partial_i \Omega=0\,.
\end{equation}
The geometry of the quaternionic K\"ahler manifold is briefly reviewed in Appendix \ref{SK} where the general properties of the quaternionic isometries $t_m$ and their description in terms of Killing vectors $k_m$ and tri-holomorphic momentum maps $\mathcal{P}_m^x$ are recalled.\par

\paragraph{Symplectically-covariant gaugings of $\mathcal{N}=2$ supergravity.}
Let us now consider the gauging of a subgroup $\mathcal{G}$ of the isometry group of the scalar manifold. The gauge generators
are conveniently written as components of an electric-magnetic vector ${\rm X}_M=({\rm X}_\Lambda,\,{\rm X}^\Lambda)$, according to the notation of \cite{deWit:2005ub}
and expanded in the generators $\{t_a,\,t_m\}$ of the isometry groups of $\mathcal{M}_{SK}$ and $\mathcal{M}_{QK}$ through the embedding tensor:
\begin{equation}
{\rm X}_M=\Theta_M{}^a\,t_a+\Theta_M{}^m\,t_m\,.
\end{equation}
The symplectic electric-magnetic duality action of ${\rm X}_M$ is described by the symplectic matrices: ${\rm X}_{MN}{}^P=\Theta_M{}^a\,t_{a\,N}{}^P $\,.
Consistency of the gauging is guaranteed by the following set of linear and quadratic constraints on the embedding tensor:
\begin{eqnarray}
&&\,\,\,\,{\rm X}_{(MNP)}\equiv {\rm X}_{(MN}{}^Q\mathbb{C}_{Q|P)}=0\,,\label{lc}\\
&&\,\,\,\,\Theta_M{}^a\Theta_N{}^bf_{ab}{}^c+{\rm X}_{MN}{}^P\,\Theta_P{}^c=0\,,\label{qc1}\\
&&\,\,\,\,\Theta_M{}^m\Theta_N{}^nf_{mn}{}^p+{\rm X}_{MN}{}^P\,\Theta_P{}^p=0\,,\label{qc2}\\
&&\,\,\,\,\Theta_M{}^a\mathbb{C}^{MN}\Theta_N{}^b=\Theta_M{}^a\mathbb{C}^{MN}\Theta_N{}^n=\Theta_M{}^m\mathbb{C}^{MN}\Theta_N{}^n=0\,.\label{qc3}
\end{eqnarray}
Conditions (\ref{qc1}), (\ref{qc2}) are closure constraints, i.e. are equivalent to
\begin{equation}
[{\rm X}_M,\,{\rm X}_N]=-{\rm X}_{MN}{}^P\,{\rm X}_P\,.\label{XXXX}
\end{equation}
The first two equalities in (\ref{qc3}) follow from (\ref{lc}) and (\ref{qc1}), (\ref{qc2}) while the last one has to be imposed independently \cite{deWit:2005ub}.
We can define gauge Killing vectors and momentum maps as follows:
\begin{equation}
k^i_M\equiv \Theta_M{}^a\,k^i_a\,,\quad k^u_M\equiv \Theta_M{}^m\,k^u_m\,,\quad\mathcal{P}_M\equiv\Theta_M{}^a\,\mathcal{P}_a\,,\quad\mathcal{P}_M^x\equiv\Theta_M{}^m\,\mathcal{P}_m^x\,.
\end{equation}
From the quadratic constraints and Eqs. (\ref{equivar1}) and (\ref{equivar2}) we find the equivariance conditions \footnote{By setting the parameter $\lambda$ of the quaternionic geometry to $\lambda =-1$.}:
\begin{eqnarray}
ig_{i\bar{\jmath}}\,k^i_{[M}\,k^{\bar{\jmath}}_{N]}=\frac{1}{2}\,{\rm X}_{MN}{}^P\,\mathcal{P}_P\,,\label{gequiv1}\\
2\,K^x_{uv}\,k^u_M\,k^v_N+\epsilon^{xyz}\,\mathcal{P}_M^y\,\mathcal{P}_N^z={\rm X}_{MN}{}^P\,\mathcal{P}_P^x\,,\label{gequiv2}
\end{eqnarray}
Using the linear constraint (\ref{lc}) on the embedding tensor we can prove the following identities:
\begin{equation}
\mathcal{P}_M\Omega^M=0\,\,,\,\,\,k_M^i\,\Omega^M=0\,.\label{newidentities}
\end{equation}
The proof is presented in Appendix \ref{ident}.\par
From (\ref{newidentities}) it also follows, as shown in Appendix \ref{ident},  that the \emph{generalized structure constants} ${\rm X}_{MN}{}^P$ are antisymmetric in the first two indices only if contracted
to the right by $\Theta_P$: ${\rm X}_{MN}{}^P\Theta_P=-{\rm X}_{NM}{}^P\Theta_P$. By virtue of this feature we find:
\begin{equation}
\overline{V}^Mk_M^i\,U_i^P\Theta_P=-{\rm X}_{MN}{}^P\,\overline{V}^M V^N\Theta_P={\rm X}_{NM}{}^P\,\overline{V}^M V^N\Theta_P=-{V}^Mk_M^{\bar{\imath}}\,\overline{U}_{\bar{\imath}}^P\Theta_P\,.\label{kMUi2}
\end{equation}
The identities (\ref{newidentities}) and (\ref{kMUi2})
 were proven in the electric case in \cite{D'Auria:1990fj}.
 Here, for the first time,  we give a general, compact proof in local special geometry of their generalization to a generic dyonic gauging, showing that they directly follow from the linear constraint on the embedding tensor.

\paragraph{The general Ward identity}

Consistency of $\mathcal{N}=2$ supergravity is based on the supersymmetry Ward identity \cite{ward}, which is required in order to cancel the  terms in the supersymmetry variation of the gauged Lagrangian, which are quadratic in the embedding tensor. It expresses a relation between the fermion
shift matrices and the scalar potential $\mathcal{V}(z,\bar{z},q)$ and has the following form:
\begin{equation}
g_{i\bar{\jmath}}\,W^{i\,AC}\overline{W}_{BC}^{\bar{\jmath}}+2\,\overline N_\alpha{}^A\,N^{\alpha}{}_{B}-12 \,\overline S^{AC}S_{BC}=\delta_A^B\,\mathcal{V}(z,\bar{z},q)\,,\label{Ward}
\end{equation}
where $W^{i\,AC},\,N^{\alpha}_{B},\,S_{AB}$ are the supersymmetry shift-matrices of the chiral gaugini $\lambda^i$, hyperini $\zeta^{\alpha}$ and gravitini $\psi_A$ respectively, $\overline{W}_{BC}^{\bar{\jmath}}= \left(W^{j\,BC}\right)^*$, $\overline N_\alpha{}^A= \left(N^\alpha_A\right)^*$, $\overline S^{AC}= \left(S_{AC}\right)^*$ being their complex conjugates:\footnote{We use the following convention for rising and lowering symplectic indices: $$v_A=\epsilon_{AB} \,v^B\,,\,\,v^A=\epsilon^{BA}\,v_B\,,\,\,\,v_\alpha=\mathbb{C}_{\alpha\beta}
\,v^\beta\,,\,\,\,v^\alpha=\mathbb{C}^{\beta\alpha}\,v_\beta\,.$$}
\begin{align}
\delta^{(\Theta)}_\epsilon\lambda^{i\text{ }A}  &  =W^{i\text{ }AB}\epsilon_{B},\label{shiftF1}\\
\delta^{(\Theta)}_\epsilon\psi_{A\text{ }\mu}  &  =iS_{AB}\gamma_{\mu}\epsilon^{B},\label{shiftF2}\\
\delta^{(\Theta)}_\epsilon\zeta^{\alpha}  &  =N_{A}^{\alpha}\epsilon^{A},\label{shiftF3}
\end{align}
where $\delta^{(\Theta)}_\epsilon$ denotes the term in the supersymmetry transformation rule of the field which is proportional to the embedding tensor.
For their definition in the electric case we refer to  \cite{D'Auria:1990fj,Andrianopoli:1996cm}.  In particular $S_{AB}$ also enters the Lagrangian as the  gravitino mass matrix whose eigenvalues on a bosonic background are the gravitino masses.
 Let us now prove the Ward identity \cite{ward} for the generic dyonic gauging of $\mathcal{N}=2$ supergravity. In this case  the fermion shifts have to be generalized to the following symplectically-invariant expressions:\footnote{Note the  relative sign between the two terms in $W^{i\,AB}$, which corrects a typo in \cite{Andrianopoli:1996cm}.}
\begin{eqnarray}
S_{AB}&=&\frac{i}{2}\,(\sigma^x)_A{}^C\epsilon_{BC}\,\mathcal{P}^x_M\,V^M\,,\label{SAB}\\
W^{i\,AB}&=&\epsilon^{AB}\,k_M^i\,\overline{V}^M-i\,(\sigma^x)_C{}^B\epsilon^{CA}\mathcal{P}^x_M\,g^{i\bar{\jmath}}
\overline{U}^M_{\bar{\jmath}}\,,\label{WiAB}\\
N_\alpha{}^A&=&2\,\mathcal{U}_{u}^A{}_\alpha\,k^u_M\,\overline{V}^M\,\,,\,\,\,\,N^\alpha{}_A\equiv (N_\alpha{}^A)^*=-2\,\,\mathcal{U}_{u\,A}{}^\alpha\,k^u_M\,{V}^M\,,\label{NAa}
\end{eqnarray}
 where  $\left(  \sigma^{x}\right)_{A}{}^{C}$ are the standard Pauli matrices.
We shall evaluate each term in the left hand side of (\ref{Ward}) separately in Appendix \ref{ident}.
Explicit calculation gives, for the left hand side of the Ward identity, the following decomposition in a singlet and a triplet of $SU(2)$:
\begin{equation}
g_{i\bar{\jmath}}\,W^{i\,AC}\overline{W}_{BC}^{\bar{\jmath}}+2\,\overline N_\alpha{}^A\,N^{\alpha}{}_{B}-12 \,\overline S^{AC}S_{BC}=\delta_B^A\,\mathcal{V}(z,\bar{z},q)+i\,Z^x\,(\sigma^x)_B{}^A\,,
\end{equation}
where
\begin{equation}
\mathcal{V}(z,\bar{z},q)=(k_M^ik_N^{\bar{\jmath}}g_{i\bar{\jmath}}+4\,h_{uv}k_M^uk_N^v)\overline{V}^M\,V^N+(U^{MN}-3\,V^M\overline{V}^N)\mathcal{P}^x_N\mathcal{P}^x_M \,, \label{potentialV}
\end{equation}
is the general symplectic invariant expression of the scalar potential given in \cite{deWit:2005ub} as a generalization to dyonic gaugings of the one given in \cite{Andrianopoli:1996cm}, and
\begin{equation}
Z^x=(-2\,
{\rm X}_{MN}{}^P\,\mathcal{P}^x_P+2\,\epsilon^{xyz}\,\mathcal{P}^y_M\mathcal{P}^z_N+4\,K^x_{uv}k^u_M\,k^v_N)\overline{V}^{M} V^{N}\,.
\end{equation}
From the equivariance condition (\ref{gequiv2}) it follow that $Z^x=0$, so that the Ward identity is proven.

\paragraph{Abelian gauging of quaternionic isometries.}
Let us now make contact with the gauging considered in this paper
which involves an abelian group of quaternionic isometries. Being only quaternionic isometries gauged, the generalized structure constants vanish: ${\rm X}_{MN}{}^P=0$, so that (\ref{gequiv2}) implies:
\begin{equation}
K^x_{uv}k^u_M\,k^v_N=-\frac{1}{2}\,\epsilon^{xyz}\,\mathcal{P}^y_M\mathcal{P}^z_N\,.
\end{equation}
Using this identity, it is easy to explicitly show  that, in this case, the three fermion-shifts all contribute to $Z^x$ and  that they cancel against one another:
 \begin{eqnarray}
 g_{i\bar{\jmath}}\,W^{i\,AC}\overline{W}_{BC}^{\bar{\jmath}}&\rightarrow &-\epsilon^{xyz}\,\mathcal{P}^y_M\mathcal{P}^z_N\overline{V}^{M} V^{N}\,,\label{w1}\\
 2\,\overline N_\alpha{}^A\,N^{\alpha}{}_{B}&\rightarrow &-2\,\epsilon^{xyz}\,\mathcal{P}^y_M\mathcal{P}^z_N\overline{V}^{M} V^{N}\,,\label{w2}\\
 -12\,\overline S^{AC}S_{BC}&\rightarrow &3\,\epsilon^{xyz}\,\mathcal{P}^y_M\mathcal{P}^z_N\overline{V}^{M} V^{N}\,.\label{w3}
 \end{eqnarray}
 We shall be interested, in what follows, in the limit of a gauged $\mathcal{N}=2$ supergravity of this kind to a rigid supersymmetric theory of $n$ vector multiplets \cite{Hughes:1986dn} (rigid limit), along the lines of \cite{FGP}. We wish here to make few general comments on the rigid limit of the Ward identity (\ref{Ward}) \cite{APT,FGP,Rocek:1997hi,Andrianopoli:2015wqa}. This will be in fact a crucial point in our analysis.

 The Ward identity of an $\mathcal{N}=2$ (abelian) rigid supersymmetric theory of $n$ vector multiplets is given by the general expression \cite{APT,FGP,Andrianopoli:2015wqa}:
 \begin{equation}
\mathring{g}_{i\bar{\jmath}}\,\mathring{W}^{i\,AC}\overline{\mathring{W}}_{BC}^{\bar{\jmath}}=\delta_B^A\,
{V}^{(APT)}_{\mathcal{N}=2}(z,\bar{z})+C_B{}^A\,,\label{wardrig}
\end{equation}
where ${V}^{(APT)}_{\mathcal{N}=2}(z,\bar{z})$ is the $\mathcal{N}=2$ scalar potential in the spontaneously broken rigid theory, which reproduces the APT one in the one-vector case, $ C_B{}^A$ is a $\mathfrak{su}(2)$-traceless matrix, $\mathring{g}_{i\bar{\jmath}}$ is the metric of the rigid special K\"ahler manifold describing the scalar fields $z^i$ in the vector multiplets and $\mathring{W}^{i\,AC}$ are the gaugini shift-matrices of the rigid theory.\par
 As shown in \cite{APT,FGP},    partial breaking of rigid supersymmetry is possible only if $ C_B{}^A\neq 0$. This happens in the presence of mutually non-local electric and magnetic Fayet-Iliopoulos terms \cite{APT}.

The symplectically-covariant relations (\ref{w1}),(\ref{w2}),(\ref{w3}) allow to clarify the meaning of the matrix $ C_B{}^A$ by relating the rigid Ward identity (\ref{wardrig}) to the supergravity one (\ref{Ward}).
To this end let us rewrite the supergravity Ward identity in the form:
\begin{equation}
g_{i\bar{\jmath}}\,W^{i\,AC}\overline{W}_{BC}^{\bar{\jmath}}=\delta_A^B\,\mathcal{V}(z,\bar{z},q)-2\,\overline N_\alpha{}^A\,N^{\alpha}{}_{B}+12 \,\overline S^{AC}S_{BC}\,,\label{Ward2}
\end{equation}
As we shall illustrate in detail in the next section, all squared fermion-shift matrices in (\ref{Ward2}) survive in the rigid limit in which the Planck mass $M_{Pl}$ is sent to infinity. In particular the left-hand-side of (\ref{Ward2}) reproduces that of  (\ref{wardrig}), while the constant matrix $C_B{}^A$ receives contribution from the terms in $N_\alpha{}^A\,N^{\alpha}{}_{B},\,S^{AC}S_{BC}$ proportional to $\sigma^x$, which are given in (\ref{w2}), (\ref{w3}). More specifically we will find that:
\begin{equation}
C_B{}^A=\lim_{M_{Pl}\rightarrow \infty}\,\frac{M_{Pl}^4}{\Lambda^4}\,\left(-i\,\epsilon^{xyz}\,\mathcal{P}^y_M\mathcal{P}^z_N\overline{V}^{M} V^{N}(\sigma^x)_B{}^A\right)\,,
\end{equation}
where $\Lambda$ is the supersymmetry-breaking scale. The same hyperini and gravitini shift-matrices also contribute terms  proportional to $\delta_B^A$ which affect the form of the scalar potential in the resulting rigid theory.
These terms are explicitly computed in (\ref{NNesp}) and (\ref{SSesp}) so that we can identify:
\begin{equation}
\mathcal{V}^{(APT)}_{\mathcal{N}=2}=\lim_{M_{Pl}\rightarrow \infty}\,\frac{M_{Pl}^4}{\Lambda^4}\,\left[\mathcal{V}(z,\bar{z},q)-(4\,h_{uv}\,k_M^u k_N^v-3\,\mathcal{P}^x_M\mathcal{P}^x_N)\overline{V}^MV^N\right]\,.
\end{equation}
As we shall prove in the next section, in the rigid limit the leading order terms in $\Theta_N{}^n V^N$ are independent of $z^i,\,\bar{z}^i$, so that:
\begin{equation}
\mathcal{V}^{(APT)}_{\mathcal{N}=2}=\lim_{M_{Pl}\rightarrow \infty}\,\frac{M_{Pl}^4}{\Lambda^4}\,\left[\mathcal{V}(z,\bar{z},q)\right]+A(q)\,.
\end{equation}
Since the fluctuations of $q^u$ are suppressed by a factor $M_{Pl}^{-1}$, see Section \ref{rigid}, in the rigid theory the hyperscalars are non-dynamical, i.e. constants. As a consequence of this, the $\mathcal{N}=2$ scalar potential of the rigid theory $\mathcal{V}^{(APT)}_{\mathcal{N}=2}$ is given by the rigid limit of the supergravity potential $\mathcal{V}$ modulo an unphysical additive constant. This was already observed in \cite{FGP} for the particular model considered there.

\section{Generalization of the APT model to $n$ vector multiplets}

 In this section we present an $\mathcal{N}=2$ supergravity model which, in the low energy  limit, gives rise to a rigid supersymmetric theory corresponding to the generalization of the APT model \cite{APT} to a generic number $n$ of vector multiplets. In particular, this procedure admits  a well defined limit to many-vector supersymmetric Born-Infeld theory.

 The minimal underlying supergravity model, considered here, consists of $\mathcal{N}=2$ supergravity
coupled to $n$ vector multiplets and a single hypermultiplet, whose scalars
parametrize the quaternionic manifold
\begin{equation}
\mathcal{M}_{QK}=\frac{SO\left(  4,1\right)  }{SO\left(  4\right)  }.
\end{equation}

Following the procedure adopted in \cite{FGP}, let us consider a special geometry  symplectic section
\begin{equation}
\Omega^{M}\left(  z^{i}\right)  =\binom{X^{\Lambda}\left(  z^{i}\right)
}{F_{\Lambda}\left(  z^{i}\right)  }\text{ \ \ \ \ \ }\Lambda=0,I,\text{
\ \ \ \ }I,i=1,\dots,n,
\end{equation}
(where $i$ are holomorphic-coordinate indices) in a symplectic frame where a holomorphic prepotential exists. Using special coordinates $z^i = \delta^i_I X^I/X^0 $, it takes the form:
\begin{equation}\label{redpre}
F\left(  X^{\Lambda}\right)  =-i\left(  X^{0}\right)  ^{2}f\left(  X^{i}%
/X^{0}\right)\ ,
\end{equation}
so that, choosing $X^0=1$:
\begin{equation}\Omega^M    =\left(
\begin{array}
[c]{c}%
1\\
z^{i}\\
-i\left(  2f-z^{i}\partial_{i}f\right) \\
-i\partial_{i}f
\end{array}
\right)  .
\end{equation}
In particular the K\"{a}hler potential becomes%
\begin{align*}
\mathcal{K}  &  =-\ln\left[  i\left(  \bar{X}^{\Lambda}F_{\Lambda}-X^{\Lambda
}\bar{F}_{\Lambda}\right)  \right]  \\
&  =-\ln\left[  2\left(  f+\bar{f}\right)  -\left(  z-\bar{z}\right)
^{i}\left(  \partial_{i}f-\overline{\partial_{i}f}\right)  \right]  .
\end{align*}

In order to generalize the procedure in  \cite{FGP} to the case of $n$ vector multiplets, we should consider a rigid limit
($ \mu=M_{Pl}/\Lambda\rightarrow\infty $, where $M_{Pl}$ denotes the Planck scale and $\Lambda$ the  supersymmetry breaking scale), leading
 to partial breaking $\mathcal{N}=2\to \mathcal{N}=1$ in a rigid supersymmetric theory.
 A crucial point, in the derivation of \cite{FGP}, was the presence of a linear term (in the holomorphic special coordinate $z$) in the expansion of the prepotential $f(z)$ in powers of $\frac 1\mu$:
  \begin{equation}
f\left(  z \right)  =\frac{1}{4}+\frac{ z }{2\mu}+\frac{\phi(z )
}{2\mu^{2}}+O\left(  \frac{1}{\mu^{3}}\right)  .
\end{equation}
In the case of many vector multiplets, we shall adopt for the holomorphic prepotential a simple generalization of the above expression which involves  a set of $n$ constant parameters $\eta_i$ and has the form
\begin{equation}
f\left(  z^{i}\right)  =\frac{1}{4}+\frac{\eta_{i}z^{i}}{2\mu}+\frac{\phi(z^i)
}{2\mu^{2}}+O\left(  \frac{1}{\mu^{3}}\right)\,  .
\label{feta}
\end{equation}
Using the standard formula for the K\"ahler potential one derives, up to order $\mu^{-3}$
\begin{align}
\mathcal{K}   =-\frac{\eta_{i}\left(  z+\bar{z}\right)  ^{i}}{\mu}-\frac{1}{\mu^{2}%
}\left[  \phi+\bar{\phi}-\left(  z-\bar{z}\right)  ^{i}\left(  \frac
{\partial_{i}\phi-\overline{\partial_{i}\phi}}{2}\right)  -\frac{\left(
\eta_{i}\left(  z+\bar{z}\right)  ^{i}\right)  ^{2}}{2}\right]  .
\label{Kahlpot}%
\end{align}
so that
\begin{align}
g_{i\bar{\jmath}}  &  =\partial_{i}\partial_{\bar{\jmath}}\mathcal{K} =\frac{1}{\mu^{2}}\mathring{g}_{i\bar{\jmath}}=\frac{1}{\mu^{2}}\left\{
\eta_{i}\eta_j-\frac{1}{2}\left(  \overline{\partial_{ij}\phi}+\partial_{ij}%
\phi\right)  \right\}  ,
\label{rigmetr}\end{align}
where $\mathring{g}_{i\bar{\jmath}}$ corresponds to the rigid special K\"{a}hler
metric. Let us note that the rigid special K\"{a}hler metric can be derived, in terms of the
(rigid) $Sp(2n)$-symplectic section
\begin{equation}\label{rigsec}
\hat{\Omega}^\mathcal{M}=\left(
\begin{array}
[c]{c}%
z^{i}\\
\partial_{i}\mathcal{F}%
\end{array}
\right) =\left(
\begin{array}
[c]{c}%
z^{i}\\
\frac i2(\eta_i\eta_j z^j -\partial_i \phi)
\end{array}
\right) \,, \quad \mathcal{M}=1,\cdots , 2n\,,
\end{equation}
from the (rigid) prepotential
\begin{equation}
\mathcal{F}=\frac{i}{4}\left[  \left(  \eta_{i}z^{i}\right)  ^{2}%
-2\phi\right]  \,. \label{Fsk}%
\end{equation}
Indeed, defining
\begin{equation}
\mathcal{F}_{i{\jmath}}=\partial_{i}\partial_{{\jmath}}\mathcal{F}= \tau_{ij}(z)=\tau_{1i{\jmath}}(z,\bar z)+i\tau_{2i{\jmath}}(z,\bar z)
\end{equation}
we find
$$
\mathring{g}_{i\bar{\jmath}}=\frac{1}{2}\tau_{2i{\bar \jmath}}(z,\bar z)
$$
where $\mathring{g}_{i\bar{\jmath}}$ is defined in equation (\ref{rigmetr}).\par
 The covariantly holomorphic symplectic section $V^{M}\equiv e^{\mathcal{K}/2}\Omega^{M}$ has the following expansion
\begin{equation}
V^{M}=\left(
\begin{array}
[c]{c}%
1-\frac{1}{2\mu}\eta_{i}\left(  z+\bar{z}\right)  ^{i}+O\left(  1/\mu
^{2}\right) \\
z^{j}-\frac{1}{2\mu}\eta_{i}\left(  z+\bar{z}\right)  ^{i}z^{j}+O\left(
1/\mu^{2}\right) \\
-i\left[  \frac{1}{2}+\frac{1}{2\mu}\left\{  \eta_{i}z^{i}-\frac{1}{2}\eta
_{i}\left(  z+\bar{z}\right)  ^{i}\right\}  \right]  +O\left(  1/\mu
^{2}\right) \\
-\frac{i}{2\mu}\eta_{j}+O\left(  1/\mu^{2}\right)
\end{array}
\right)  . \label{V}%
\end{equation}
In this framework, the physical meaning of the constant parameters $\eta_i$ appearing in the  symplectic  section $\hat{\Omega}^\mathcal{M}$ and in the metric $\mathring{g}_{i\bar{\jmath}}$ of the rigid theory needs to be clarified. We will see in Section \ref{eta} that a natural interpretation of $\eta_i$ can be given in supergravity, as charges associated with the gauging procedure, by performing a different choice of symplectic frame.

Postponing this issue to next section, let us consider, for the time being, a gauging of two translational isometries in the hypermultiplet sector involving both electric and magnetic charges \cite{Dall'Agata:2003yr,D'Auria:2004yi}. This gauging
can be described in terms of a (redundant) symplectic vector of gauge generators ${\rm X}_M\equiv ({\rm X}_\Lambda,\,{\rm X}^\Lambda)$, expressed
as linear combinations of the isometry generators $t_m$, $m=1,\dots,\dim\mathcal{G}$, of the quaternionic K\"ahler manifold through an embedding tensor \cite{deWit:2002vt,deWit:2005ub}:
\begin{equation}
{\rm X}_M=\Theta_M{}^m\,t_m\,.
\end{equation}
We choose the gauging only to involve two translational isometries $t_m$ ($m=1,2$)  and the embedding tensor to depend on constant charges $e,\sigma,m^i$ as follows
\begin{equation}
\Theta_{M}^{\text{ \ }m}=\left(  \Theta_{M}^{\text{ \ }1},\Theta
_{M}^{\text{ \ }2}\right)  =\left(
\begin{array}
[c]{cc}%
\Theta_{0}^{\text{ \ }1} & \Theta_{0}^{\text{ \ }2}\\
\Theta_{i}^{\text{ \ }1} & \Theta_{i}^{\text{ \ }2}\\
\Theta^{0\text{ }1} & \Theta^{0\text{ }2}\\
\Theta^{i\text{ }1} & \Theta^{i\text{ }2}%
\end{array}
\right)  =\left(
\begin{array}
[c]{cc}%
e/\mu^{2} & \sigma/\mu^{2}\\
0 & 0\\
0 & 0\\
m^{i}/\mu  & 0
\end{array}
\right)\,,  \label{embed}
\end{equation}
satisfying the locality condition
\begin{equation}\label{loc}
\mathbb{C}
^{MN}\Theta_{M}^{\text{ \ }m}\Theta_{N}^{\text{ \ }n}=0\,.
\end{equation}

The embedded Killing vectors
$k_{M}^{\text{ \ }u}=\left(  k_{\Lambda}^{\text{ \ }u},k^{\Lambda\text{ }%
u}\right)$ are related to the geometrical ones $k_{m}^{\text{ \ }u}$ ($m=1,\dots,\dim\mathcal{G}$) generating the isometry group $\mathcal{G}$ of $\mathcal{M}_{QK}$ by:
\begin{equation}
k_{M}^{\text{ \ }u}=\Theta_{M}^{\text{ \ }m}k_{m}^{\text{ \ }u}.
\end{equation}

The fermion shifts $\delta^{(\Theta)}_\epsilon$, entering the supersymmetry transformation laws (\ref{shiftF1})-(\ref{shiftF3}) of the fermion fields, are written in terms of the embedding tensor in a symplectic covariant way in (\ref{SAB})-(\ref{NAa}). To obtain their explicit form for the  $\mathcal{N}=2$ gauged supergravity under consideration, we should set $k_M^i=0$, since our gauging does not involve special K\"ahler isometries.\par
Denoting by $\varphi$ and $\vec{q}\equiv \{q^1,\,q^2,\,q^3\}$ the four hyper-scalars in the solvable parametrization,  the metric of the quaternionic K\"ahler manifold has the form:
\begin{equation}
ds^2=\frac{1}{2}\,\left(d\varphi^2+e^{2\varphi}\,d\vec{q}\cdot d\vec{q}\right)\,,\label{metricqk}
\end{equation}
and the corresponding vielbein  $\mathcal{U}_{A|u}^{\alpha}$  reads \cite{FGP}:
\begin{equation}
\mathcal{U}_{A}^{\alpha }=\mathcal{U}_{A|u}^{\alpha }dq^{u}=-\frac{1}{2}%
\epsilon ^{\alpha \beta }\left[ d\varphi \, \delta_{\beta A}+i\,e^{\varphi }d\vec{q}\cdot \vec{%
\sigma}_\beta{}^A\right]\,.\label{quat}
\end{equation}
The metric (\ref{metricqk}) is invariant under constant translation of the three axions: $\vec{q}\rightarrow \vec{q}+\vec{c}$. We choose to gauge the two translations $t_n$ acting on $q^2,\,q^3$.
The quaternionic momentum maps $\mathcal{P}_{m}^{x}$  associated with translational isometries have the general form:\footnote{For homogeneous quaternionic K\"ahler manifolds this relation holds only for those isometries whose action on the coset representative does not imply a compensating transformation in the isotropy group, see Appendix \ref{SK} for a general proof. These include translational isometries.}
 \begin{equation}
\mathcal{P}_{m
}^{x}=  -k_{m}^{\text{ \ }u} \omega^x_u\,,
\end{equation}
where $\omega^x_u$ denotes the $SU(2)$-connection on $\mathcal{M}_{QK}$.
For the gauging under consideration (\ref{embed}) which involves the two  traslational isometries $t_n$, the momentum maps can be explicitly computed to be%
\[
\mathcal{P}_{m}^{x}=\left(  \mathcal{P}_{1}^{x},\mathcal{P}_{2}%
^{x}\right) = \delta^x_m e^{\varphi} ,
\]
with%
\begin{align}
\mathcal{P}_{1}^{x}  &  =\left(  0,1,0\right)  e^{\varphi},\\
\mathcal{P}_{2}^{x}  &  =\left(  0,0,1\right)  e^{\varphi}.
\end{align}
Later, in Section \ref{eta}, the two hyperscalars $q^2,q^3$ will be dualized into antisymmetric tensor fields $B_{n|\,\mu\nu}$.

\subsection{The rigid limit and partial supersymmetry breaking}\label{scales}

The partial supersymmetry breaking is recovered considering the limit $\mu
=\frac{M_{Pl}}{\Lambda}\rightarrow\infty$. We will follow here the prescription in \cite{FGP}. Later, in Section \ref{rigid}, we will consider the low energy limit of the Lagrangian starting from a different, $\mu$-dependent, symplectic frame of the supergravity theory where the rigid limit of the symplectic structure is more transparent, and which will require a different rescaling of the physical fields.  To explicitly perform the limit on the fermionic shifts  (which are written in  natural units $c= \hbar= M_{Pl}=1$)  we will first reintroduce the appropriate dependence on the Planck scale $M_{Pl}$ and on the supersymmetry breaking scale $\Lambda$, due to the gauging, in the supergravity expressions.
Taking into account that the scale $\Lambda$ is related to the gravitino mass via $\Lambda^2 = M_{Pl}\,m_{\frac32}$, and that the Special-K\"ahler sigma-model metric rescales according to (\ref{rigmetr}), the canonically normalized kinetic terms are recovered  by the rescaling \cite{FGP}:
 \begin{align}
x^{\mu}  &  \rightarrow M_{Pl}x^{\mu}\text{, \ \ \ \ \ \ \ \ }\epsilon
\rightarrow M_{Pl}^{1/2}\epsilon\text{, \ \ \ \ \ \ \ \ \ \ \ \ \ \ \ \ }%
\nonumber\\
\psi_{\mu}  &  \rightarrow M_{Pl}^{-3/2}\psi_{\mu}\text{, \ \ \ \ }%
\lambda\rightarrow\left(  M_{Pl}\Lambda^{2}\right)  ^{-1/2}\lambda\text{,
\ \ \ \ }\zeta^{\alpha}\rightarrow M_{Pl}^{-3/2}\zeta^{\alpha}\text{.}%
\nonumber\\
&  \label{resc}%
\end{align}
Using the rescaling of eq. $\left(  \ref{resc}\right)  $ we find that in the rigid
limit the shifts of the fermions read%
\begin{align}
\delta\lambda^{i\text{ }A} &  =i\Lambda^{2}\epsilon^{CA}\left[  \mathring
{g}^{i\bar{\jmath}}\left(  e_{\bar{\jmath}}^{x}-\tau_{1\bar{\jmath}%
k}m^{k\text{ }x}\right)  +\frac{i}{2}m^{i\text{ }x}\right]  \left(  \sigma
^{x}\right)  _{C}^{\text{ \ }B}e^{\varphi}\epsilon_{B},\nonumber\\
\delta\psi_{A\text{ }\mu} &  =-\frac{\Lambda^{2}}{2}\epsilon_{BC}\left[
e^{x}-i\frac{\eta_{j}}{2}m^{j\text{ }x}\right]  \left(  \sigma^{x}\right)
_{A}^{\text{ \ }C}e^{\varphi}\gamma_\mu\epsilon^{B},\nonumber\\
\delta\zeta^{\alpha} &  =i\Lambda^{2}\epsilon^{\alpha\beta
}\left[  e^{x}-i\frac{\eta_{j}}{2}m^{j\text{ }x}\right]  \left(  \sigma
^{x}\right)^{\alpha}{}_{A}e^{\varphi}\epsilon^{A},\label{shifts}%
\end{align}
where we have used the following definitions%
\begin{align}
e^{x} &  =\left(  0,e,\sigma\right) =(0,e^m)\, ,\nonumber\\
m^{i\text{ }x} &  =\left(  0,m^{i},0\right)  =(0,m^{im})\,,\label{charges}\\
e_{i}^{x} &  =\eta_{i}e^{x}. \nonumber
\end{align}
As we will see in detail by the analysis of the lagrangian in the rigid limit in  Section \ref{rigid}, the hypermultiplets decouple in the rigid theory so that $\varphi$ becomes a constant and $\delta\lambda^{iA}$ get the characteristic form of the gaugino shifts in a rigid theory in the presence of electric-magnetic Fayet-Iliopoulos parameters $\mathbb{P}^{x\mathcal{M}}= \left(m^{ix},e^x_i\right)$. and the momentum maps $\mathcal{P}^{xM}$ yield constant  Fayet-Iliopoulos (FI) parameters $\mathbb{P}^{x\mathcal{M}}= \left(m^{ix},e^x_i\right)$.
The precis relation between the momentum maps $\mathcal{P}^{xM}$ and the FI terms can be directly read from the gaugino shift:
\begin{equation}
 \mathring g^{i \jmath}\bar U^{M}_{\bar\jmath} \mathcal{P}^{x}_{M}=\frac{e^\varphi}{\mu}\,\left[  \mathring
{g}^{i\bar{\jmath}}\left(  e_{\bar{\jmath}}^{x}-\tau_{1\bar{\jmath}%
k}m^{k\text{ }x}\right)  +\frac{i}{2}m^{i\text{ }x}\right] =\frac{e^\varphi}{\mu}\,\mathring
{g}^{i\bar{\jmath}} \bar U^{\mathcal{M} }_{\bar\jmath}\mathbb{P}^x_\mathcal{M}\,,\label{PFI}
\end{equation}
where $U^{\mathcal{M} }_i$ are related to the rigid symplectic sections introduced in (\ref{rigsec}) by
$U^{\mathcal{M} }_i= \partial_i \hat\Omega^\mathcal{M}$. We emphasize here that in this formulation of the rigid limit, the FI terms are expressed not only in terms of the parameters $e,\,\sigma,\,m^i$ defining the embedding tensor (the gauging parameters), but also in terms of the parameters $\eta_i$ characterizing the special geometry through the choice of the prepotential (\ref{feta}). We shall discuss in the next Section a different formulation in which the FI terms fully descend from the supergravity gauging parameters codified in the embedding tensor.

For the case of one vector multiplet, $n=1$, eq. (\ref{shifts}) reproduces the results of \cite{FGP} leading to the APT model.\par
\paragraph{Partial supersymmetry breaking.}
Applying the general discussion at the end of Sect. \ref{gauge}, we find that the gaugino shifts in the rigid theory satisfy the rigid Ward identities (\ref{wardrig}) where \cite{Andrianopoli:2015wqa}:\footnote{Recall that in the rigid special K\"ahler geormetry the matrix $\mathcal{M}$ is defined by the relation
$$U^{\mathcal{M}\mathcal{N}}= \partial_i \hat \Omega^\mathcal{M}\partial_{\bar\jmath} \hat {\bar\Omega}{}^\mathcal{N} \mathring g^{i\bar\jmath}= \frac 12\left(\mathcal M^{\mathcal{M}\mathcal{N}} - \mathbb C^{\mathcal{M}\mathcal{N}}\right)\,,$$
and is positive definite.}
\begin{align}
V^{(APT)}_{\mathcal{N}=2}&=\frac{e^{2\varphi}}{2}\,\mathcal{M}(z,\bar{z})^{\mathcal{M}\mathcal{N}}\,\mathbb{P}^x_\mathcal{M}\mathbb{P}^x_\mathcal{N}\,,\nonumber\\
C_B{}^A&=e^{2\varphi}\,\xi^x\,(\sigma^x)_B{}^A\,,\,\,\,\,\,\xi^x=\frac{1}{2}\epsilon^{xyz}\,
\mathbb{P}^y_\mathcal{M}\mathbb{P}^z_\mathcal{N}\mathbb{C}^{\mathcal{M}\mathcal{N}}=
\epsilon^{xyz} {m}^{y i}  {e}_i^{\,z}\,.\label{Cxi}
\end{align}
In the rigid theory, as explained earlier, the hyperscalars are non-dynamical constants. In particular the factor $e^{2\varphi}$ can be absorbed in a redefinition of the FI terms. For this reason we shall neglect it in the discussion below.\par
Partial supersymmetry breaking \cite{Cecotti:1985sf,Cecotti:1984wn,Hughes:1986dn,Hughes:1986fa,Ferrara:1995gu,APT,Partouche:1996yp,FGP,Fre:1996js} in the rigid theory requires $\delta_\epsilon \lambda^{iA}$ to vanish along a suitable direction in the supersymmetry parameter space. This in turn implies that the $2\times 2$ matrix on  the left hand side of  (\ref{wardrig}) should have, on the vacuum defined by $z_0^i,\,\bar{z}^{\bar{\imath}}_0$, one zero eigenvalue. As explained in \cite{Andrianopoli:2015wqa}, this condition can be cast in the following symplectic invariant form for the scalar potential:
\begin{equation}
V^{(APT)}_{\mathcal{N}=2}(z_0,\,\bar{z}_0)=\sqrt{I_4}\,,\label{Vxi}
\end{equation}
where $I_4\equiv \sum_{x=1}^3\xi^x\xi^x$ is a quartic symplectic invariant defined in terms of the FI parameters. Being $V^{(APT)}_{\mathcal{N}=2}$ positive definite, we can have partial supersymmetry breaking only if $I_4\neq 0$, that is if $\xi^x=\epsilon^{xyz} {m}^{yi}{e}_i^{\,z}\neq 0$, in which case Eq. (\ref{Vxi}) would fix $z_0^i,\,\bar{z}^{\bar{\imath}}_0$ in terms of the FI parameters. In this case the effective $\mathcal{N}=1$ potential is
$$
V^{(APT)}_{\mathcal{N}=1}(z,\,\bar{z})\equiv V^{(APT)}_{\mathcal{N}=2}(z,\,\bar{z})-\sqrt{I_4}\,,
$$
and the infra-red dynamics is captured by a multi-filed Born-Infeld action, as shown in \cite{Ferrara:2014oka}. If $\xi^x=0$, condition (\ref{Vxi}) could only be satisfied if $\mathbb{P}_{\mathcal{M}}^x=0$ or at the boundary of the moduli space, in which case the vacuum would preserve the full $\mathcal{N}=2$ supersymmetry. A non-vanishing matrix $C_A{}^B$, or equivalently $\xi^x$, is therefore a crucial ingredient in order to have partial supersymmetry breaking in the rigid theory, thus evading previously stated no-go theorems \cite{Witten:1982df,Cecotti:1984rk}.\par
Notice that partial supersymmetry breaking in the parent supergravity theory is a more stringent condition: On a bosonic Minkowski vacuum it can occur only if
the supersymmetry transformations of \emph{all}  the fermionic fields vanish along a same spinorial direction $\epsilon^A$.  Since the eigenvalues of $S_{AC}\overline S^{BC}$ (which is proportional to $N_A^\alpha\,N_\alpha^B$) are:
\begin{equation}
\lambda_\pm=\frac{e^{2\varphi}}{4}\left[e^2+\left(\sigma\pm \frac{\eta_i m^i}{2}\right)^2\right]\,,
\end{equation}
partial supersymmetry breaking in the hidden sector (defined by the gravitational multiplet and the hypermultiplet) can occur only if $m^{i\,x},\,e_i^x$ in (\ref{charges}) are not generic but satisfy the condition:
\begin{equation}
e=0\,\,;\,\,\,\,\eta_i m^i=\pm 2\,\sigma\,.
\end{equation}
Therefore for generic $m^{i\,x},\,e_i^x$, provided $\xi^x\neq 0$, we can have partial supersymmetry breaking in the visible sector \emph{albeit all sypersymmetry is broken in the hidden one}. An analogous phenomenon was observed in \cite{FGP} in  one vector multiplet case.\par
As a final remark, the same multi-vector, ${\rm U}(1)^n$-rigid supersymmetric theory could be obtained from an $\mathcal{N}=2$ supergravity with a more general  quaternionic K\"ahler manifold, including the vast class of manifolds in the image of the \emph{c-map} \cite{Cecotti:1988qn}. In the latter case, the gauging should involve abelian generators in the universal Heisenberg algebra of isometries of these manifolds \cite{heisenberg,Louis:2009xd}.

\section{\bigskip Interpretation of the constant parameters $\eta_i$ as charges}\label{eta}
As we have recalled in the previous Section, partial supersymmetry breaking in rigid supersymmetry crucially requires the quantity $\xi^x$ in (\ref{Cxi}) to be different from zero
\begin{equation}\label{nonloc2}
\xi^x\equiv \frac 12 \epsilon^{xyz}\mathbb{P}^{y\mathcal{M}}\mathbb{P}^{z\mathcal{N}}\mathbb{C}_{\mathcal{MN}} = \epsilon^{xyz} {m}^{yi}{e}_i^{\,z}\neq 0\,,
\end{equation}
where  $e^y_i, m^{zi}$ are given by (\ref{charges}).
This relation looks like a non-locality condition. However, the choice of embedding tensor (\ref{embed}) implies that the locality condition
\begin{equation}
\Theta^{m}_\mathcal{M}\Theta^{n}_\mathcal{N} \mathbb{C}^{\mathcal{MN}}=2\Theta^{I[m }\Theta^{n]}_I =0\,,
 \end{equation}
 is satisfied in the rigid theory so that, recalling the definition of the momentum maps $\mathcal P^{x}_{\mathcal{M}}= \mathcal P^x_m \Theta_{\mathcal{M}}^m$, the condition $\epsilon^{xyz}\mathcal{P}^{y \mathcal{M}}\mathcal{P}^{z \mathcal{N}}\mathcal{C}_{\mathcal{MN}}=0$  is satisfied in the chosen frame. This is not in contradiction with (\ref{nonloc2}) since the Fayet-Iliopoulos parameters $\mathbb{P}_\mathcal{M}^x$ of the rigid theory \emph{are not} the simple restriction of the supergravity momentum maps to the ${\rm Sp}(2n,\mathbb{R})$-index $\mathcal{M}$, but $\mathcal{P}_M^x$ and $\mathbb{P}_\mathcal{M}^x$  are rather related through (\ref{PFI}), which non-trivially involves the contribution from the index $0$ of the symplectic section, keeping memory of the graviphoton.
Moreover, as emphasized earlier,  Eqs. (\ref{rigmetr}) and (\ref{rigsec}) show that the geometry of the rigid theory \emph{in the chosen coordinate frame } depends in a non-trivial way on the constant parameters $\eta_i$,  also appearing in (\ref{nonloc2}) through the charges $e^y_i= e^y \eta_i$.

As we are going to see, the embedding of the theory in supergravity  allows to clarify the topological role of all the constant  parameters involved in the gauging, showing that  the $\eta_i$ required in the special geometry of the rigid theory in order to implement partial supersymmetry breaking (with its BI low-energy limit) can be traded with   charges via a symplectic rotation involving a redefinition of the special coordinates \emph{in the underlying supergravity theory}.
\\
Indeed, let us consider the (electric) symplectic transformation in supergravity:
\begin{equation}
S(\eta,\mu)=\begin{pmatrix} 1&\eta_i/\mu &0&0\\
0&\mathbf{\frac 1\mu 1_n}&0&0\\
0&0& 1&0\\
0&0& -\eta_i & \mu \mathbf{1_n}\end{pmatrix} \label{seta}
\end{equation}
inducing the following rotation in the symplectic section (\ref{V}):
\begin{equation}
\tilde \Omega = S\cdot \Omega=
\left(\begin{array}{c}
      X^0 + \frac 1\mu\eta_i X^i\\
      \frac 1\mu X^i \\
      F_0 \\
      \mu F_i - \eta_i F_0
    \end{array}\right)= \left(\begin{array}{c}
      \tilde X^0  \\
     \tilde  X^i \\
     \tilde  F_0 \\
     \tilde  F_i
    \end{array}\right)\,.
\end{equation}
The new holomorphic prepotential is $\tilde F(\tilde X)= F(X)$. Since the new special coordinates $\tilde z^i$ are related to the old ones by
\begin{equation}
\tilde z^i = \frac{z^i}{\mu + \eta_j z^j}= \frac 1\mu \omega^i\,,
  \end{equation}
  then the reduced prepotential $\tilde f(\tilde z)$ is related to $f(z)$ by (see (\ref{redpre})):
\begin{eqnarray}
\tilde f(\tilde z)&=& (1+\frac 1\mu \eta_j z^j)^{-2}f(z)\end{eqnarray}
that is
\begin{equation}\label{tilde f}
\tilde f(\tilde z)=\left( \frac 14 + \frac 1{2\mu^2}\tilde \phi(\tilde z) + O(\frac 1{\mu^3})\right)
\end{equation}
where $\tilde \phi(\tilde z)$ is related to $\phi(z)$ by
$\tilde \phi(\tilde z) = \phi(z) - \frac 12 (\eta_i  z^i)^2\equiv \Phi(\omega)$.
We note that in the new frame \emph{the contribution linear in $\tilde z$ has disappeared from} (\ref{tilde f}) (to be compared with (\ref{feta}).). Moreover, after the symplectic rotation, the   covariantly holomorphic symplectic sections $\tilde V^M= e^{\frac{\mathcal{K}}2}\tilde \Omega^M$ and $\tilde U^M_i= D_i \tilde V^M$ can be written in a generic coordinate frame with holomorphic coordinates $\omega^i$ and behave, in the rigid limit $\mu \to \infty$, as:
\begin{eqnarray}\label{vring}
\tilde V^{M}(\omega)&=& \left(
\begin{array}
[c]{c}%
X^0 \\
0 \\
F_0   \\
0
\end{array}
\right)\,+ \frac 1\mu \left(
\begin{array}
[c]{c}%
0 \\
 \mathring{X}^I (\omega) \\
0  \\
\mathring{F}_I (\omega)
\end{array}
\right) \,+O\left(  1/\mu
^{2}\right) \,;
\\\label{uring}
\tilde U^{M}_i(\omega)&=&\frac 1\mu\left(
\begin{array}
[c]{c}%
0 \\
\partial_i   \mathring{X}^I  \\
0   \\
\partial_i   \mathring{F}_I
\end{array}
\right) \,+O\left(  1/\mu
^{2}\right) \,,
\end{eqnarray}
where $\mathring\Omega^\mathcal{M}\equiv (\mathring X^I, \mathring F_I)$ ($I=1,\cdots n$) denotes the symplectic section or the rigid theory
(in special coordinates $\mathring X^I(\omega) = \omega^i$, $ \mathring F_I (\omega)= \frac{\partial   \Phi}{\partial \omega^i}$).
We observe that in the new frame the symplectic structure $Sp(2n+2)$ of the supergravity theory flows in the rigid limit to a manifest $Sp(2n)$ structure. In particular,  the $0$-directions have a different $\mu$-rescaling  with respect to the $\mathcal{M}$-directions. They are then directly associated with the Hodge-bundle of the local special geometry (that is to the graviphoton direction) which is projected-out in the low energy limit. Still, the special-geometry sigma-model metric in supergravity  is related to its counterpart $\mathring g_{i\bar\jmath}$ in the rigid limit by:
\begin{equation}
\label{gg0}
g_{i\bar\jmath} = \frac 1{\mu^2}\mathring g_{i\bar\jmath}\,,
\end{equation}
while the relations of special geometry imply a  low-energy rescaling of the   vector-kinetic-matrix $\mathcal{N}_{\Lambda\Sigma}$ corresponding to the following identification of the matrix $\mathring{\mathcal{N}}_{\Lambda\Sigma}$ of the rigid theory:
\begin{equation}
\label{N0}
\mathcal{N}_{00} =  \mathring{\mathcal{N}}_{00}\,, \quad \mathcal{N}_{IJ} =  \mathring{\mathcal{N}}_{IJ}\,,\quad
\mathcal{N}_{0I} = \frac 1{\mu } \mathring{\mathcal{N}}_{0I}\,.
\end{equation}
\vskip 5mm

The symplectic transformation (\ref{seta}) also acts on the embedding tensor (\ref{embed}) as
\begin{equation} \label{embednew}
\tilde\Theta^{m}_M = \Theta^{m}_N \cdot (S^{-1})^{N}_{\ M} =
\frac 1{\mu^2}\left( e^m , - \eta_i  e^m, \eta_i m^{im},   m^{im}\right)\,= \,\frac 1{\mu^2}\mathring \Theta^{m}_M \,,
\end{equation}
where we have introduced the tensor $\mathring \Theta^{m}_M $, whose components in the $\mathcal M$ directions will define the FI parameters  of  the rigid theory.

In the new frame the parameters $\eta_i$ play the role of charges, since  $\tilde \Theta_i^m =\eta_i e^m$ are the  electric charges associated with the vector multiplets and $\tilde \Theta^{0m}= \eta_i m^{im}$ are the magnetic charges associated with the graviphoton.
Note that in the old frame both of them were zero.

As a consequence, the new embedding tensor (\ref{embednew}) of the supergravity theory obeys the same locality condition (\ref{loc}) as the old one, but now
\begin{equation}
\tilde \Theta^{\Lambda [m}\tilde\Theta_\Lambda^{n]} =0\quad \Rightarrow \quad \tilde \Theta^{0[m} \tilde \Theta_0^{n]}= -\tilde \Theta^{i[m} \tilde \Theta_i^{n]}=\frac 1{\mu^4} \eta_i m^{i[m}e^{n]}\neq 0\,.\label{nonlocsusy}
\end{equation}
Furthermore, as already observed, in the new frame the graviphoton is identified with the 0 direction of the vector field strengths, which is not true in the old frame; we will explicitly show this in the next section, see in particular eq. (\ref{graf}). Since in the rigid limit the graviphoton decouples  from the spectrum, we find that the rigid supersymmetric theory found as low-energy limit of supergravity  in the new frame exhibits a \emph{non-locality in superspace}, which means that, as we are going to discuss in the following, the non-locality only affects the fermionic directions of superspace, while \emph{it does not emerge as a non-locality on space-time.}  This clarifies the meaning of (\ref{nonloc2}), as expressing indeed the non locality of the rigid theory, when all the constant parameters needed for the partial breaking of supersymmetry  are expressed as electric and magnetic charges in the embedding tensor. In what follows, for the sake of notational simplicity, we shall denote the embedding tensor $\tilde \Theta$ in the new frame simply by $\Theta$.\par
 Let us analyze the effects of the non-locality (\ref{nonlocsusy}), which is intimately related to the supersymmetric structure of the theory:
 \begin{itemize}
   \item Since the superspace non-locality of the rigid theory is related to the non-triviality of the fiber bundle associated with the graviphoton in the rigid limit, the supergravity modes associated with the underlying $\mathcal{N}=2$ supergravity theory (the gravitini   and hyperini, together with their bosonic partners) still freely propagate in the rigid theory (see (\ref{shifts})) even if decoupled from the visible sector, as already observed in \cite{FGP}.
This justifies the presence of the $SU(2)$-Lie algebra valued term $C_A{}^B$ in the supersymmetry Ward-identity of the spontaneously broken rigid theory, which is understood as the contribution to the rigid Ward identity from gravitini and hyperini, as explicitly shown in Sect. \ref{gauge}.
   \item It is known  \cite{Dall'Agata:2003yr,Sommovigo:2004vj,deWit:2005ub,Andrianopoli:2007ep} that, in the presence of magnetic charges $m^{\Lambda n}$ in supersymmetric theories, the natural symplectic frame to deal with them is  rotated with respect to the purely electric one,
allowing for the presence of
 antisymmetric tensors $B_{n|\,\mu\nu}$, coupled to the gauge fields $A^\Lambda$ in the combinations
 $\hat F^\Lambda_{\mu\nu}=F^\Lambda_{\mu\nu}+ 2m^{\Lambda n}B_{n\mu\nu}$ and realizing the so-called anti-Higgs mechanism for the gauge fields. \footnote{The fermionic shifts, found in \cite{FGP} and generalized to $n$ vector multiplets in section \ref{scales} of the present paper, are in fact naturally recovered in the symplectic frame where some of the hyper-scalars are dualized to tensor fields, as one can explicitly check by comparison with Section 3 of \cite{Dall'Agata:2003yr}, and in particular eqs. (3.13) - (3.15) there.}
 The $\mathcal{N}=2$ supersymmetric Free Differential Algebra in four dimensions contains in particular, in the case where the antisymmetric tensors dualize scalars in the quaternionic sector\footnote{In \cite{D'Auria:2001kv} the index $I$ was used for our index $n$, to label the quaternionic scalars to be dualized into antisymmetric tensors. Moreover the corresponding field strengths were defined as:
 $$H^{(3)}_n= dB_n-\omega_{n\,A}{}^B\, \bar{\psi}^A\wedge\gamma_a \psi_B \wedge V^a\,,$$
 where $\omega_{n\,A}{}^B \equiv \frac{i}{2}\,\omega^x_u\,k^u_n\,(\sigma^x)_A{}^B$. \\Taking into account that $\mathcal P^x_n=-\omega^x_u\,k^u_n$, and that here $k^u_n=\delta^u_n$, the definition (\ref{fda2}) follows.}
 \begin{eqnarray}
   \hat F^{(2)\Lambda} &\equiv & d A^\Lambda + 2m^{\Lambda n}B_n +( L^\Lambda(z) \bar{\psi}_A\wedge\psi_B\,\,\epsilon^{AB}+ h.c.) \label{fda1}\\
   H^{(3)}_n &\equiv & dB_n +\frac{i}{2}\,\mathcal P^x_n\,(\sigma^{x})_A{}^B \bar{\psi}^A\wedge\gamma_a \psi_B \wedge V^a\label{fda2}
 \end{eqnarray}
 where $L^\Lambda$ are the upper-part of the special geometry symplectic sections $V^M$ and $\mathcal P^x_n$ are functions of the hyperscalars \cite{D'Auria:2004yi}. From (\ref{fda1}) and (\ref{fda2}) we get that the closure of the free differential algebra requires
 \begin{equation}\label{nonlocf}
   d\hat F^\Lambda =  \Theta^{\Lambda n}\left(2 H_n -i\mathcal P^x_n\,(\sigma^{x})_A{}^B  \bar{\psi}^A\wedge\gamma_a \psi_B \wedge V^a\right)\,,
 \end{equation}
 where we have identified $m^{\Lambda n}$ with $\Theta^{\Lambda\,n}$.
  As discussed above,  in the low energy limit the hyperscalars are not suppressed but tend
   to constants, in such a way that $\Theta_M{}^n\mathcal P^x_n(q)$ become constants $ \Theta_M{}^n\mathbb{P}^x_n \neq 0$ whose restriction to the non-zero indices $\Theta_{\mathcal{M}}{}^n\mathbb{P}^x_n$ yield the FI parameters. Then, from eq. (\ref{nonlocf}), taking into account the decoupling of the tensor fields, the closure of the supersymmetric free differential algebra gives
  \begin{equation}
   d\hat F^I \propto i \Theta^{I m}\mathbb{P}^x_m \,(\sigma^{x})_A{}^B  \bar{\psi}^A\wedge\gamma_a \psi_B\wedge V^a+\dots\neq 0\,.\label{nonlocvecrig}
 \end{equation}
  As previously discussed this equation is the superspace counterpart of the fact that on space-time the commutator of two supersymmetries acts on the gauge field $A_\mu^I$ as a gauge transformation proportional to the magnetic FI parameters, as stressed in reference \cite{Ferrara:1995gu}.\footnote{Recall that, according to (\ref{embednew}), $\Theta^{I m}=\mathring{\Theta}^{I m}/\mu^2$, so that one would expect that the right hand side of (\ref{nonlocvecrig}) vanish in the rigid limit. However, in the same limit, the leading component of $\psi_A$ along the fermionic directions is $M_{Pl}\,d\theta_A$, so that
  $$\Theta^{I m}\mathbb{P}^x_m \,(\sigma^{x})_A{}^B  \bar{\psi}^A\wedge\gamma_a \psi_B\wedge V^a\rightarrow \mathring{\Theta}^{I m}\mathbb{P}^x_m \,(\sigma^{x})_A{}^B  \bar{d\theta}^A\wedge\gamma_a d\theta_B\wedge V^a\neq 0\,.$$

  }
 \end{itemize}

\section{\bigskip The rigid limit of the N=2 Lagrangian}\label{rigid}

In this section we want to recover the rigid limit of the $\mathcal{N}=2$ supergravity lagrangian corresponding to partial breaking of supersymmetry, and whose gauge structure has been discussed in the previous section.

We will work in the symplectic frame defined in section \ref{eta}, where the gauging structure of the theory is unveiled and shown to involve the presence of magnetic charges (and where it is not necessary to rely on a particular choice of coordinates in the special-geometry sigma-model). According to this, the natural framework to perform the limit is the version of the lagrangian where some of the scalars of the hypermultiplets are Hodge-dualized to antisymmetric tensors $B_{m|\mu\nu}$ \cite{Dall'Agata:2003yr,D'Auria:2004yi,Sommovigo:2004vj,deWit:2005ub, Andrianopoli:2007ep}. We will then refer to the lagrangian in \cite{D'Auria:2004yi}.

In order to perform the rigid limit, it is convenient to  reintroduce in the lagrangian, which is usually written in natural units $c= \hbar=1$, but with also $M_{Pl}=1$, the appropriate scale dimensions, as anticipated in Section \ref{scales}.
This will be performed in two steps: We will first explicitly write the correct Planck-mass dependence of the physical fields in the supergravity lagrangian and then, after considering the low energy ($\mu\to \infty$) behavior of the special-geometry sigma-model sector, we will get the appropriate redefinitions of the  physical fields appearing in the rigid supersymmetric theory.

\begin{itemize}
\item
The canonical scale dimensions of the fields of the theory in natural units $c= \hbar=1$ are:
\begin{eqnarray*}
&&[dx^\mu]=M^{-1}\,, \quad [\partial_\mu]= M\,,\quad [d\theta^A]=[\epsilon^A]=M^{-\frac{1}{2}}\,, \\
&&\quad[A_\mu^\Lambda]= [B_{\mu\nu}^x]=M\,,\quad [z_{\small{(can.)}}^i]=
[q_{\small{(can.)}}^u]=M\,\,,\,\,\,[\psi^A_\mu]= [\lambda^A]= [\zeta^\alpha]= M^{3/2}\,,
\end{eqnarray*}
while the  embedding tensor is  dimensionless.
Since the scalars $z^i,q^u$ appear in the theory through non-linear sigma-models, we will keep them  dimensionless (that is we will consider $z^i \equiv z^i_{\small{(can.)}}/M_{Pl}$, $q^u \equiv q^u_{\small{(can.)}}/M_{Pl}$).

According with this prescription, the supergravity lagrangian can be organized in terms of Planck-scale powers and reads, up to four fermions terms:
\begin{eqnarray}\label{lmasses}
\mathcal{L}=\mathcal{L}_{(4)}+\mathcal{L}_{(2)}+\mathcal{L}_{(1)}+\mathcal{L}_{(0)}+\mathcal{L}_{(-1)}
\end{eqnarray}
where
\begin{eqnarray}
\mathcal{L}_{(4)}&=&M_{Pl}^4 \mathcal{V}(z,q)\label{pot}\\
\mathcal{L}_{(2)}&=&M_{Pl}^2  \left(-\frac{R}{2}+g_{i\bar{\jmath}}\partial^{\mu}%
z^{i}\partial_{\mu}\bar{z}^{\bar{\jmath}}+h_{uv}\partial_{\mu}q^{u}\partial^{\mu}q^{v}\right)\label{L2}\\
\mathcal{L}_{(1)}&=&M_{Pl}\left\{\frac{\epsilon^{\mu\nu\rho\sigma}}{\sqrt{-g}}  \left[2\mathcal{H}_{m|\nu\rho\sigma} A_{u}^{m}\partial_{\mu}q^{u}-{2} {B_{m|\mu\nu}}%
 \Theta_{\Lambda}^{\text{ }m}\left(  \hat{\mathcal{F}}_{\rho\sigma}^{\Lambda}-M_{Pl}  \Theta^{\text{ }\Lambda n}B_{n|\rho\sigma}\right)
\right]\right.+\nonumber\\
&&{\qquad}+\left(2S_{AB}\bar{\psi}_{\mu}^{A}\gamma^{\mu\nu}\psi_{\nu
}^{B}+ig_{i\bar{\jmath}}W^{iAB}\bar{\lambda}_{A}^{\bar{\jmath}}\gamma_{\mu
}\psi_{B}^{\mu}+2iN_{\alpha}^{A}\bar{\zeta}^{\alpha}\gamma_{\mu}\psi_{A}^{\mu
}\right.\nonumber\\
&&{\qquad}\left.\left.  +\mathcal{M}^{\alpha\beta}\bar{\zeta}_{\alpha}\zeta_{\beta}+\mathcal{M}%
_{iB}^{\alpha}\bar{\zeta}_{\alpha}\lambda^{iB}+\mathcal{M}_{iAjB}\bar{\lambda
}^{iA}\lambda^{jB}+\text{h.c.}\right)\right\}\label{L1}\\
\mathcal{L}_{(0)}&=& i\left(  \mathcal{\bar{N}}%
_{\Lambda\Sigma}\hat{\mathcal{F}}_{\mu\nu}^{-\Lambda}\hat{\mathcal{F}}^{-\Sigma\mu\nu
}-\mathcal{N}_{\Lambda\Sigma}\hat{\mathcal{F}}_{\mu\nu}^{+\Lambda}\hat{\mathcal{F}}
^{+\Sigma\mu\nu}\right)+6M^{mn}\mathcal{H}_{m\mu\nu\rho}\mathcal{H}_{n}^{\text{ }\mu
\nu\rho} +\nonumber\\
&&+\frac{\epsilon^{\mu\nu\lambda\sigma}}{\sqrt{-g}
}\left(  \bar{\psi}_{\mu}^{A}\gamma_{\nu}\rho_{A|\lambda\sigma}-\bar{\psi
}_{A|\mu}\gamma_{\nu}\rho_{\lambda\sigma}^{A}\right)  -\frac{i}{2}
g_{i\bar{\jmath}}\left(  \bar{\lambda}^{iA}\gamma^{\mu}\nabla_{\mu}\lambda
_{A}^{\bar{\jmath}}+\bar{\lambda}_{A}^{\bar{\jmath}}\gamma^{\mu}\nabla_{\mu
}\lambda^{iA}\right) +\nonumber\\
&&  -i\left(  \bar{\zeta}^{\alpha}\gamma^{\mu}\nabla_{\mu}\zeta_{\alpha}%
+\bar{\zeta}_{\alpha}\gamma^{\mu}\nabla_{\mu}\zeta^{\alpha}\right)+\nonumber\\
&& -g_{i\bar{\jmath}}\partial_{\mu}\bar{z}^{\bar{\jmath
}}\left(\bar{\psi}_{A}^{\mu}\lambda^{iA}-\bar{\lambda}^{iA}\gamma^{\mu\nu}\psi_{A\nu}+h.c.\right)  -2\mathcal{U}_{u}^{\alpha A}\partial_{\mu}q^{u}\left(\bar{\psi}_{A}^{\mu}%
\zeta_{\alpha}-\bar{\zeta
}_{\alpha}\gamma^{\mu\nu}\psi_{A\nu} +h.c.\right)\nonumber\\
&& \label{L0}\\
\mathcal{L}_{(-1)}&=& M_{Pl}^{-1}\Bigl\{\hat{\mathcal{F}}_{\mu\nu}^{-\Lambda}I_{\Lambda\Sigma}\left[  L^{\Sigma}
\bar{\psi}^{A\mu}\psi^{B\nu}\epsilon_{AB}-4i\bar{f}_{\bar{\imath}}^{\Sigma
}\bar{\lambda}_{A}^{\bar{\imath}}\gamma^{\nu}\psi_{B}^{\mu}\epsilon
^{AB}  +\frac{1}{2}\nabla_{i}f_{j}^{\Sigma}\bar{\lambda}^{iA}\gamma
^{\mu\nu}\lambda^{jB}\epsilon_{AB}+\right.\nonumber\\
&&{\qquad}\left.-L^{\Sigma}\bar{\zeta}_{\alpha}\gamma
^{\mu\nu}\zeta_{\beta}
\mathbb{C}
^{\alpha\beta}\right]+h.c.+\nonumber\\
&&{\qquad}
+ 2\mathcal{M}^{mn}\mathcal{H}_{m}^{\text{ }\mu\nu\rho}\left[\mathcal{U}%
_{n}^{\text{ }A\alpha}\left(3i \bar{\psi}_{A\mu}\gamma_{\nu\rho}\zeta_{\alpha}
+\bar{\psi
}_{A\mu}\zeta_{\alpha}\right)+ i\Delta_{n\alpha}^{\text{
\ }\beta\text{\ }}\zeta_{\beta}\gamma_{\mu\nu\rho}\zeta^{\alpha}\right] \Bigr\}\,,\label{L-1}
\end{eqnarray}
where $h_{uv}$, $A^{m}_u$, $M^{mn}$ are the components of the quaternionic metric after dualizition of  the scalars $q^m$ to antisymmetric tensors $B_{m|\mu\nu}$, $\hat{\mathcal{F}}_{\mu\nu}^{\Lambda
}:= {\mathcal{F}}^{\Lambda}_{\mu
\nu }+{2}M_{Pl}\, \Theta^{\Lambda m}B_{\mu
\nu m}$ are the gauge field-strengths undergoing the anti-Higgs mechanism introduced in (\ref{fda1}) (in our case $ \Theta^{\Lambda m}=m^{\Lambda m}=\frac 1{\mu^2}\eta_i m^{i m}$),
 $\mathcal{F}_{\mu\nu}^{\pm\Lambda}=\frac{1}{2}\left(
\mathcal{F}_{\mu\nu}^{\Lambda}\pm\frac{i}{2}\epsilon_{\mu\nu\rho\sigma
}\mathcal{F}^{\Lambda\rho\sigma}\right)  $ denotes projection on (anti)self-dual part.
For the definition of the  mass-matrices we refer to \cite{Andrianopoli:1996cm} and \cite{D'Auria:2004yi}. We will present their symplectic-covariant generalization, together with their relation with the quantities appearing in the rigid theory, in eqs. (\ref{m}),(\ref{mi}),(\ref{mij}) below.

\item
To perform the rigid limit $\frac{M_{Pl}}\Lambda \equiv \mu \to \infty$ of the lagrangian, where $\Lambda$ denotes the scale of supersymmetry breaking  defining the gauging, we should first consider the limit of the kinetic terms for the various fields which should appear in the rigid lagrangian. This will define the relation between supergravity fields and their rigid counterparts. We will generally identify the fields of the rigid supersymmetric theory with an upper ring, to distinguish them from the supergravity fields.

According to the discussion in Section \ref{eta}, the special-K\"ahler metric rescales, for $\mu \to \infty$, as (\ref{gg0}), so that the kinetic terms of scalars and spinors in the vector multiplets in the rigid limit read (from (\ref{L2}) and (\ref{L0}):
$$\frac 1{\mu^2}\mathring g_{i\bar{\jmath}}\left[M_{Pl}^2\partial^{\mu}%
z^{i}\partial_{\mu}\bar{z}^{\bar{\jmath}} -\frac{i}{2}
\left(  \bar{\lambda}^{iA}\gamma^{\mu}\nabla_{\mu}\lambda
_{A}^{\bar{\jmath}}+\bar{\lambda}_{A}^{\bar{\jmath}}\gamma^{\mu}\nabla_{\mu
}\lambda^{iA}\right)\right]
$$
This implies that the  gaugini and of the rigid theory should be related to their supegravity relatives as:
\begin{equation}
\mathring \lambda^{iA}= \frac 1\mu \lambda^{iA}
\end{equation}
while  the holomorphic scalars should not be rescaled ($\mathring z^i=z^i$), so that
$$\mathcal L_{rig}= \cdots \mathring g_{i\bar{\jmath}}\left[\Lambda^2\partial^{\mu}%
\mathring z^{i}\partial_{\mu}\bar{ \mathring
z}^{\bar{\jmath}} -\frac{i}{2}
\left( \bar{\mathring{ \lambda}}^{iA}\gamma^{\mu}\nabla_{\mu}\mathring\lambda
_{A}^{\bar{\jmath}}+\bar{\mathring{ \lambda}}_{A}^{\bar{\jmath}}\gamma^{\mu}\nabla_{\mu
}\mathring\lambda^{iA}\right)\right]+\cdots
$$
Furthermore, the components of the  gauge kinetic matrix $\mathcal N_{\Lambda\Sigma}$ rescale as (\ref{N0})
 so that  the gauge kinetic term reads, at low energies:
$$I_{\Lambda\Sigma }F^\Lambda_{\mu\nu} F^{\Sigma|\mu\nu}=\mathring I_{00} F^0_{\mu\nu}  F^{0|\mu\nu}+ \mathring I_{IJ} F^I_{\mu\nu}  F^{J|\mu\nu} + \frac 2\mu \mathring I_{0I}  F^0_{\mu\nu}  F^{I|\mu\nu}+ \mathcal O(1/\mu^2)$$
where we defined $I_{\Lambda\Sigma}\equiv Im(\mathcal N_{\Lambda\Sigma})$.
This implies that no redefinition of the gauge vectors should should be applied:
\begin{equation}
\mathring A^\Lambda_\mu =A^\Lambda_\mu\,,\label{aring}
\end{equation}
and that the interaction term between $F^0$ and $F^I$ goes to zero in the limit.
Given (\ref{vring}), (\ref{uring}), (\ref{embednew}) and (\ref{aring}), we can then identify the low energy limit of the self-dual components of the graviphoton $T^-_{\mu\nu}$ and of the matter vectors $G^{-i}_{\mu\nu}$. We find:
\begin{eqnarray}
T^-_{\mu\nu}&\equiv & I_{\Lambda\Sigma}L^\Lambda   F^{-\Sigma}_{\mu\nu}\to \mathring I_{00}\mathring X^0 \mathring F^{-0}_{\mu\nu} +   O(\frac 1{\mu })\label{graf}
\\
g_{i\bar\jmath}G^{-i}_{\mu\nu}&\equiv &
\frac i2 I_{\Lambda\Sigma}f_{ \bar\jmath}^\Lambda   F^{-\Sigma}_{\mu\nu}\to
 \frac i{2\mu}   \mathring I_{IJ}\mathring f_i^I \mathring F^{-J}_{\mu\nu} + O(\frac 1{\mu^2})\label{matf}\,,
\end{eqnarray}
showing that, in the rigid limit, the gauge-index 0 corresponds to the graviphoton direction, while the gauge-index $I$ to the matter-vectors directions.

The rescalings of the fermion shifts and spinor mass matrices follow from the low energy limit of the symplectic sections and embedding tensor discussed in section \ref{eta}. They are:\footnote{The matrices (\ref{w})-(\ref{mij}) are related to one another by differential ``gradient-flow'' equations \cite{D'Auria:2001kv}.}

\begin{align}
W^{i\text{ }AB}  &
=\frac 1\mu \mathring W^{i\text{ }AB}\,, \label{w}
\\
S_{AB}  &
=\frac 1{\mu^2}\mathring S_{AB} \,,\label{s}
\\
N_{\text{ }A}^{\alpha}  &
= \frac 1{\mu^2}\mathring N_{\text{ }A}^{\alpha} \, ,\label{n1}
\\
\mathcal{M}^{\alpha\beta}  &  =-\mathcal{U}_{u}^{\alpha A}
\mathcal{U}_v^{\beta B}\epsilon_{AB} \Theta_{M}^{\ m}
\nabla^{[ u}k_{m}^{v]}V^{M}=
\frac 1{\mu^2}
\mathring{\mathcal{M}}^{\alpha\beta} \,,\label{m}
\\
\mathcal{M}_{iB}^{\alpha}  &  =-4\mathcal{U}_{Bu}^{\alpha}%
 \Theta_{M}^{\text{ \ }m}k_{m}^{u}U_{i}^{\text{ }M}=\frac 1{\mu^3}\mathring{\mathcal{M}}_{iB}^{\alpha}   \,,\label{mi}
\\
\mathcal{M}_{iAjB}  &  =\left(\sigma_{x}\epsilon^{-1}\right)_{AB} \Theta_{M}^{\text{ \ }m}\mathcal{P}_{m}^{x}\nabla_{j} U_{i}^{\text{ }M}
= \frac 1{\mu^3}\mathring{\mathcal{M}}_{iAjB}\,
.\label{mij}
\end{align}
Consequently, the scalar potential rescales, for $\mu \to\infty$, as
$\mathcal{V}= \frac 1{\mu^4}\mathring{\mathcal{V}}\,.$

The various contributions to the lagrangian (\ref{lmasses}), when written in terms of the rescaled fields, read:
\begin{eqnarray}
\mathcal{L}_{(4)}&=&\Lambda^4  \mathring{\mathcal{V}}(z,q)\label{potring}\\
\mathcal{L}_{(2)}&=&M_{Pl}^2  \left(-\frac{R}{2}+h_{uv}\partial_{\mu}q^{u}\partial^{\mu}q^{v}\right)+ \Lambda^2 \mathring g_{i\bar{\jmath}}\partial^{\mu}%
\mathring z^{i}\partial_{\mu} \mathring{\bar{z}}^{\bar{\jmath}}\label{L2mu}\\
\mathcal{L}_{(1)}&=&
M_{Pl}\left\{ \frac{\epsilon^{\mu\nu\rho\sigma}}{\sqrt{-g}}   \left[2\mathcal{H}_{m|\nu\rho\sigma} A_{u}^{m}\partial_{\mu}q^{u}-\frac{2}{ \mu^2} {B_{m|\mu\nu}}%
\mathring \Theta_{\Lambda}^{\text{ }m}\left(  \hat{\mathcal{F}}_{\rho\sigma}^{\Lambda}-\frac{M_{Pl}}{\mu^2}  \mathring \Theta^{\text{ }\Lambda n}B_{n|\rho\sigma}\right)
\right]\right.+\nonumber\\
&&{\qquad}+\frac 1{\mu^2}\left(2\mathring S_{AB}\bar{\psi}_{\mu}^{A}\gamma^{\mu\nu}\psi_{\nu
}^{B}+i\mathring g_{i\bar{\jmath}}\mathring W^{iAB}\mathring {\bar{\lambda}}_{A}^{\bar{\jmath}}\gamma_{\mu
}\psi_{B}^{\mu}+2i\mathring N_{\alpha}^{A}\bar{\zeta}^{\alpha}\gamma_{\mu}\psi_{A}^{\mu
}+\text{h.c.}\right) +
\nonumber\\
&&{\qquad} \left.+\frac 1{\mu^2}\left(\mathring{\mathcal{M}}^{\alpha\beta}\bar{\zeta}_{\alpha}\zeta_{\beta}+\mathring{\mathcal{M}}
_{iB}^{\alpha}\bar{\zeta}_{\alpha}\mathring \lambda^{iB}+\text{h.c.}\right)\right\}+\Lambda\left(\mathring{\mathcal{M}}_{iAjB}\mathring {\bar{\lambda}}^{iA}\mathring \lambda^{jB}+\text{h.c.}\right).
\label{L1mu}\\
\mathcal{L}_{(0)}&=& i\left(  \mathcal{\bar{N}}%
_{\Lambda\Sigma}\hat{\mathcal{F}}_{\mu\nu}^{-\Lambda}\hat{\mathcal{F}}^{-\Sigma\mu\nu
}-\mathcal{N}_{\Lambda\Sigma}\hat{\mathcal{F}}_{\mu\nu}^{+\Lambda}\hat{\mathcal{F}}
^{+\Sigma\mu\nu}\right)+6M^{mn}\mathcal{H}_{m |\mu\nu\rho}\mathcal{H}_{n}^{\text{ }\mu
\nu\rho} +\nonumber\\
&&+\frac{\epsilon^{\mu\nu\lambda\sigma}}{\sqrt{-g}
}\left(  \bar{\psi}_{\mu}^{A}\gamma_{\nu}\rho_{A|\lambda\sigma}-\bar{\psi
}_{A|\mu}\gamma_{\nu}\rho_{\lambda\sigma}^{A}\right)  -\frac{i}{2}
\mathring g_{i\bar{\jmath}}\left(  \mathring{\bar{\lambda}}^{iA}\gamma^{\mu}\nabla_{\mu}\mathring \lambda
_{A}^{\bar{\jmath}}+\mathring{\bar{\lambda}}_{A}^{\bar{\jmath}}\gamma^{\mu}\nabla_{\mu
}\mathring \lambda^{iA}\right) +\nonumber\\
&&  -i\left(  \bar{\zeta}^{\alpha}\gamma^{\mu}\nabla_{\mu}\zeta_{\alpha}%
+\bar{\zeta}_{\alpha}\gamma^{\mu}\nabla_{\mu}\zeta^{\alpha}\right)+\nonumber\\
&& -\frac 1\mu\mathring g_{i\bar{\jmath}}[\partial_{\mu}{\bar{z}}^{\bar{\jmath
}}\left(\bar{\psi}_{A}^{\mu}\mathring \lambda^{iA}-\mathring{\bar{\lambda}}^{iA}\gamma^{\mu\nu}\psi_{A\nu}\right)+h.c. ] -2\mathcal{U}_{u}^{\alpha A}\partial_{\mu}q^{u}\left(\bar{\psi}_{A}^{\mu}%
\zeta_{\alpha}-\bar{\zeta
}_{\alpha}\gamma^{\mu\nu}\psi_{A\nu} +h.c.\right)\nonumber\\
&& \label{L0mu}\\
\mathcal{L}_{(-1)}&=&\Lambda^{-1} {{\mathcal{F}}}_{\mu\nu}^{-I}\mathring I_{IJ}\Bigl[ \frac{1}{2}\nabla_{i}\mathring f_{j}^{J}\mathring{\bar{\lambda}}^{iA}\gamma
^{\mu\nu}\mathring \lambda^{jB}\epsilon_{AB}+ h.c. \Bigr]+\nonumber\\
&&+ M_{Pl}^{-1}\Bigl\{{\mathcal{F}}_{\mu\nu}^{-0}\mathring I_{00}\mathring L^{0}\left[
\bar{\psi}^{A\mu}\psi^{B\nu}\epsilon_{AB} -\bar{\zeta}_{\alpha}\gamma
^{\mu\nu}\zeta_{\beta}
\mathbb{C}
^{\alpha\beta}+h.c.\right]+\nonumber\\
&&
\qquad - {\mathcal{F}}_{\mu\nu}^{-I}\mathring I_{IJ}\Bigl[{4i\mathring {\bar{f}}_{\bar{\imath}}^{J
}\mathring {\bar{\lambda}}_{A}^{\bar{\imath}}\gamma^{\nu}\psi_{B}^{\mu}\epsilon
^{AB}+ h.c.} \Bigr]\nonumber\\
&&
\qquad + 2\mathcal{M}^{mn}\mathcal{H}_{m}^{\text{ }\mu\nu\rho}\left[\mathcal{U}%
_{n}^{\text{ }A\alpha}\left(3i \bar{\psi}_{A\mu}\gamma_{\nu\rho}\zeta_{\alpha}
+\bar{\psi
}_{A\mu}\zeta_{\alpha}\right)+ i\Delta_{n\alpha}^{\text{
\ }\beta\text{\ }}\zeta_{\beta}\gamma_{\mu\nu\rho}\zeta^{\alpha}\right] \Bigr\}\,,\label{L-1mu}
\end{eqnarray}
and it reduces, in the limit $\mu\to \infty$, to:
\begin{eqnarray}
\mathcal{L}_{(4)}&=&\Lambda^4  \mathring{\mathcal{V}}(z,q)\label{potringlim}\\
\mathcal{L}_{(2)}&=&M_{Pl}^2  \left(-\frac{R}{2}+h_{uv}\partial_{\mu}q^{u}\partial^{\mu}q^{v}\right)+ \Lambda^2 \mathring g_{i\bar{\jmath}}\partial^{\mu}%
  z^{i}\partial_{\mu} {\bar{z}}^{\bar{\jmath}}\label{L2mulim}\\
\mathcal{L}_{(1)}&=&
2\frac{\epsilon^{\mu\nu\rho\sigma}}{\sqrt{-g}}M_{Pl}\mathcal{H}_{m |\nu\rho\sigma} A_{u}^{m}\partial_{\mu}q^{u}+\Lambda\left(\mathring{\mathcal{M}}_{iAjB}\mathring{\bar{\lambda
}}^{iA}\mathring \lambda^{jB}+\text{h.c.}\right).
\label{L1mulim}\\
\mathcal{L}_{(0)}&=& i\left(  \mathring{\mathcal{\bar{N}}}
_{\Lambda\Sigma} {\mathcal{F}}_{\mu\nu}^{-\Lambda} {\mathcal{F}}^{-\Sigma\mu\nu
}-\mathring {\mathcal{N}}_{\Lambda\Sigma} {\mathcal{F}}_{\mu\nu}^{+\Lambda} {\mathcal{F}}
^{+\Sigma\mu\nu}\right)+6M^{mn}\mathcal{H}_{m \mu\nu\rho}\mathcal{H}_{n}^{\text{ }\mu
\nu\rho} +\nonumber\\
&&+\frac{\epsilon^{\mu\nu\lambda\sigma}}{\sqrt{-g}
}\left(  \bar{\psi}_{\mu}^{A}\gamma_{\nu}\rho_{A|\lambda\sigma}-\bar{\psi
}_{A|\mu}\gamma_{\nu}\rho_{\lambda\sigma}^{A}\right)  -\frac{i}{2}
\mathring g_{i\bar{\jmath}}\left(  \mathring{\bar{\lambda}}^{iA}\gamma^{\mu}\nabla_{\mu}\mathring \lambda
_{A}^{\bar{\jmath}}+\mathring{\bar{\lambda}}_{A}^{\bar{\jmath}}\gamma^{\mu}\nabla_{\mu
}\mathring \lambda^{iA}\right) +\nonumber\\
&&  -i\left(  \bar{\zeta}^{\alpha}\gamma^{\mu}\nabla_{\mu}\zeta_{\alpha}%
+\bar{\zeta}_{\alpha}\gamma^{\mu}\nabla_{\mu}\zeta^{\alpha}\right) -2\mathcal{U}_{u}^{\alpha A}\partial_{\mu}q^{u}\left(\bar{\psi}_{A}^{\mu}%
\zeta_{\alpha}-\bar{\zeta
}_{\alpha}\gamma^{\mu\nu}\psi_{A\nu} +h.c.\right)\nonumber\\
&& \label{L0mulim}\\
\mathcal{L}_{(-1)}&=&\Lambda^{-1}{{\mathcal{F}}}_{\mu\nu}^{-I}\mathring I_{IJ}\Bigl[ \frac{1}{2}\nabla_{i}\mathring f_{j}^{J}\mathring{\bar{\lambda}}^{iA}\gamma
^{\mu\nu}\mathring \lambda^{jB}\epsilon_{AB}+ h.c. \Bigr]\,.\label{L-1mulim}
\end{eqnarray}
Note that the supergravity lagrangian reduces to an observable sector corresponding to the rigid lagrangian of \cite{APT}, undergoing spontaneous breaking to $\mathcal{N}=1$ supersymmetry, plus an hidden sector, fully decoupled from the observable sector:
\begin{eqnarray}
\mathcal{L}_{sugra}\to \mathcal{L}_{APT}+\mathcal{L}_{hidden}
\end{eqnarray}
where \footnote{As observed in Sect. \ref{gauge},  the scalar potential of the APT-model differs from $\mathring{\mathcal{V} }$ for an additive term, function of the hyperscalars only.}
\begin{eqnarray}
 \mathcal{L}_{APT}&=& \Lambda^2 \mathring g_{i\bar{\jmath}}\partial^{\mu}
  z^{i}\partial_{\mu} {\bar{z}}^{\bar{\jmath}}
 -\frac{i}{2}
\mathring g_{i\bar{\jmath}}\left(  \mathring{\bar{\lambda}}^{iA}\gamma^{\mu}\nabla_{\mu}\mathring \lambda
_{A}^{\bar{\jmath}}+\mathring{\bar{\lambda}}_{A}^{\bar{\jmath}}\gamma^{\mu}\nabla_{\mu
}\mathring \lambda^{iA}\right)+\nonumber\\
&& +i\left(  \mathring{\mathcal{\bar{N}}}
_{IJ} {\mathcal{F}}_{\mu\nu}^{-I} {\mathcal{F}}^{-J\mu\nu
}-\mathring {\mathcal{N}}_{IJ} {\mathcal{F}}_{\mu\nu}^{+I} {\mathcal{F}}
^{+J\mu\nu}\right)+\nonumber\\
&&
  +\Lambda^4  \mathring{\mathcal{V}}
 +\Lambda\left(\mathring{\mathcal{M}}_{iAjB}\mathring{\bar{\lambda
}}^{iA}\mathring \lambda^{jB}+\text{h.c.}\right) +\nonumber\\
&&
  +\Lambda^{-1} {{\mathcal{F}}}_{\mu\nu}^{-I}\mathring I_{IJ}\Bigl[ \frac{1}{2}\nabla_{i}\mathring f_{j}^{J}\mathring{\bar{\lambda}}^{iA}\gamma
^{\mu\nu}\mathring \lambda^{jB}\epsilon_{AB} + h.c. \Bigr]
 \\
 \mathcal{L}_{hidden}&=&M_{Pl}^2  \left(-\frac{R}{2}+h_{uv}\partial_{\mu}q^{u}\partial^{\mu}q^{v}\right)+i\left(  \mathring{\mathcal{\bar{N}}}
_{00} {\mathcal{F}}_{\mu\nu}^{-0} {\mathcal{F}}^{-0\mu\nu
}-\mathring {\mathcal{N}}_{00} {\mathcal{F}}_{\mu\nu}^{+0} {\mathcal{F}}
^{+0\mu\nu}\right)+\nonumber\\
&&+6M^{mn}\mathcal{H}_{m|\mu\nu\rho}\mathcal{H}_{n}^{\text{ }\mu
\nu\rho} +2\frac{\epsilon^{\mu\nu\rho\sigma}}{\sqrt{-g}}M_{Pl}\mathcal{H}_{m|\nu\rho\sigma} A_{u}^{m}\partial_{\mu}q^{u}+\nonumber\\
&&+\frac{\epsilon^{\mu\nu\lambda\sigma}}{\sqrt{-g}
}\left(  \bar{\psi}_{\mu}^{A}\gamma_{\nu}\rho_{A|\lambda\sigma}-\bar{\psi
}_{A|\mu}\gamma_{\nu}\rho_{\lambda\sigma}^{A}\right)   -i\left(  \bar{\zeta}^{\alpha}\gamma^{\mu}\nabla_{\mu}\zeta_{\alpha}%
+\bar{\zeta}_{\alpha}\gamma^{\mu}\nabla_{\mu}\zeta^{\alpha}\right) +\nonumber\\
&& -2\mathcal{U}_{u}^{\alpha A}\partial_{\mu}q^{u}\left(\bar{\psi}_{A}^{\mu}%
\zeta_{\alpha}-\bar{\zeta
}_{\alpha}\gamma^{\mu\nu}\psi_{A\nu} +h.c.\right)\label{hidden}
\end{eqnarray}
Note that  in the low energy limit the space-time metric, the graviphoton, the  antisymmetric tensors  and  the scalars of the hypermultiplet sector, together with their fermionic super partners obey the field equations of free waves not interacting with the rest. In particular,  the  metric should be chosen as a constant background and the hyperscalars  set to constant values.

\end{itemize}

\section{Conclusions and Outlook}
In this paper we have investigated the  supergravity origin of a ${\rm U}(1)^n$, rigid, partially-broken $\mathcal{N}=2$ supersymmetric theory whose infra-red limit is described by the multi-field BI action of \cite{Ferrara:2014oka}. The high-energy supergravity is characterized by a \emph{visible sector} described by the $n$ vector multiplets surviving the rigid limit, and by a \emph{hidden} one consisting of the gravitational multiplet and by a hypermultiplet, which decouple as the Planck mass is sent to infinity. This model also features a dyonic gauging of two translational quaternionic isometries which, for suitable choices of the embedding tensor, allows for a spontaneous partial supersymmetry breaking. In this parent gauged supergravity  we have devised a symplectic frame in which the electric and magnetic FI terms of the resulting rigid theory directly descend from the embedding tensor defining the dyonic gauging. The mutual non-locality
of the electric and magnetic FI terms, which is essential for the partial breaking of rigid $\mathcal{N}=2$ supersymmetry, is shown to be related, by the locality condition on the supergravity embedding tensor, to a the simultaneous presence of both electric and magnetic charges for the graviphoton.\par
It would be interesting to extend this analysis to allow for the presence of hypermultiplets in the rigid model. An other direction of further investigation would be the extension of the rigid limit studied in the present work to spontanously broken $\mathcal{N}>2$ supergravities which could allow to derive from them, in a suitable limit, the multi-field BI theory of \cite{Ferrara:2014oka}.

\section*{Acknowledgements}
M.T. wishes to thank Thomas Ortin for inspiring discussions. \  Two of the authors (P.C., E.R.) were supported by
grants from the Comisi\'{o}n Nacional de Investigaci\'{o}n Cient\'{\i}fica y
Tecnol\'{o}gica (CONICYT) and from the Universidad de Concepci\'{o}n, Chile.
\ P.C. and E.R. were supported in part by FONDECYT Grants N${{}^\circ}$ 1130653.

\appendix
\section{Special K\"ahler and Quaternionic K\"ahler Manifolds}\label{SK}
\paragraph{Special K\"ahler Manifolds}
A special K\"ahler manifold \cite{Strominger:1990pd,D'Auria:1990fj,Ceresole:1995jg,Andrianopoli:1996cm} $\mathcal{M}_{SK}$ is a Hodge- K\"ahler manifold endowed with  a flat, symplectic, holomorphic bundle  satisfying certain defining properties. If $\Omega(z)=(\Omega^M(z))$ denotes a section of the  holomorphic bundle, $M=1,\dots, 2n+2$, in some local trivialization:
\begin{equation}
\Omega(z)=\begin{pmatrix}X^\Lambda(z)\cr F_\Lambda(z)\end{pmatrix}\,\,,\,\,\,\Lambda=0,\dots, n\,,
\end{equation}
then in the same patch the K\"ahler potential reads:
\begin{equation}
\mathcal{K}(z,\bar{z})=-\log[i\,\overline{\Omega}(\bar{z})^T\mathbb{C}\Omega(z)]\,,\label{Komapp}
\end{equation}
where $\mathbb{C}=(\mathbb{C}_{MN})$ is the ${\rm Sp}(2(n+1),\mathbb{R})$-invariant matrix;
\begin{equation}
\mathbb{C}\equiv \begin{pmatrix}{\bf 0} & {\bf 1}\cr -{\bf 1} & {\bf 0}\end{pmatrix}\,.
\end{equation}
As in all K\"ahler manifolds the metric has the form:
\begin{equation}
g_{i\bar{\jmath}}=\partial_i\partial_{\bar{\jmath}}\mathcal{K}\,,\label{kmetric}
\end{equation}
so that the K\"ahler 2-form
\begin{equation}
K\equiv i\,g_{i\bar{\jmath}}\,dz^i\wedge d\bar{z}^{\bar{\jmath}}\,,
\end{equation}
is closed: $dK=0$  so that, in the given patch,
\begin{equation}\label{dq}
   K=dQ
\end{equation}
where   $Q$ is the $U(1)$ K\"ahler connection 1-form
\begin{equation}\label{connec}
  Q= -\frac {i}{2}\left[\partial_i  \mathcal{K} \,dz^i- c.c.\right]
\end{equation}

The transition functions connecting overlapping coordinate patches $U_{({\tt m})},\,U_{({\tt n})}$ on $\mathcal{M}_{SK}$, act on $\Omega(z)$ as follows:
\begin{equation}
\Omega_{({\tt m})}=e^{f_{({\tt m,n})}}\,\mathbb{M}^{-T}_{({\tt m,n})}\,\Omega_{({\tt n})}\,,\label{fOm}
\end{equation}
where $f_{({\tt m,n})}=f_{({\tt m,n})}(z)$ is a holomorphic function and $\mathbb{M}_{({\tt m,n})}$ is a constant ${\rm Sp}(2(n+1),\mathbb{R})$ matrix. The corresponding action on $\mathcal{K}$ amounts to a K\"ahler transformation:
\begin{equation}
\mathcal{K}_{({\tt m})}=\mathcal{K}_{({\tt n})}-f_{({\tt m,n})}-\bar{f}_{({\tt m,n})}\,.  \label{fK}
\end{equation}
We can define a \emph{covariantly holomorphic} section $V(z,\bar{z})$ as follows:
 \begin{equation}
 V(z,\bar{z})=(V^M(z,\bar{z}))=\begin{pmatrix}L^\Lambda\cr M_\Lambda\end{pmatrix}\equiv e^{\frac{\mathcal{K}}{2}}\,\Omega(z)\,.\label{defV}
 \end{equation}
 The action of the transition functions on $V$ amount to a constant symplectic transformation combined with a ${\rm U}(1)$-phase related to the K\"ahler transformation:
 \begin{equation}
V_{({\tt m})}=e^{i\,{\rm Im}(f_{({\tt m,n})})}\,\mathbb{M}^{-T}_{({\tt m,n})}\,V_{({\tt n})}\,,\label{fV}
\end{equation}
We define the following ${\rm U}(1)$-covariant derivatives on $V$:
\begin{equation}
U_i=D_i V\equiv \left(\partial_i+\frac{\partial_i\mathcal{K}}{2}\right)V\,\,,\,\,\,\bar{D}_{\bar{\imath}} V\equiv \left(\partial_{\bar{\imath}}-\frac{\partial_{\bar{\imath}}\mathcal{K}}{2}\right)V=0\,,
\end{equation}
the last equality follows from the definition (\ref{defV}) of $V$ and implies that $V$ is \emph{covariantly holomorphic}.
From the definition of $V$ and (\ref{Komapp}) it follows that $V^T\mathbb{C} \overline{V}=i$.\par
In a special K\"ahler manifold the section $V$ and its covariant derivative $U_i$ need to satisfy the following properties:
\begin{equation}
D_iU_j\equiv \partial_iU_j+\frac{\partial_i\mathcal{K}}{2}\,U_j-\Gamma_{ij}^k\,U_k=i\,C_{ijk}\,g^{k\bar{k}}\,\overline{U}_{\bar{k}}\,\,,\,\,\,
D_i\overline{U}_{\bar{\jmath}}=g_{i\bar{\jmath}}\,\overline{V}\,,\,\,\,V^T\mathbb{C}U_i=0\,,\,\,\,V^T\mathbb{C}\overline{U}_{\bar{k}}=0\,,\label{defre}
\end{equation}
the last equality being a consequence of $V^T\mathbb{C} \overline{V}=i$.\par
Using $V$ and its covariant derivatives, we can construct the following matrix:
\begin{equation}
\mathbb{L}(z,\bar{z})^M{}_{\underline{N}}\equiv(V^M,\bar{{\tt e}}_{\bar{I}}{}^{\bar{\imath}}\overline{U}^M_{\bar{\imath}},\,\overline{V}^M,\,{{\tt  e}}_I{}^{i}U_i^M)\,,
\end{equation}
where ${{\tt  e}}_I{}^{i}$ are the inverse vielbein matrices $g_{i\bar{\jmath}}=\sum_{I=\bar{I}=1}^n{{\tt  e}}_i{}^{I}\bar{{\tt  e}}_{\bar{\jmath}}{}^{\bar{I}}$, and $\underline{N}$ is a holonomy group index. Eqs. (\ref{defre}) imply the following property of $\mathbb{L}$ \cite{Andrianopoli:1996ve}:
\begin{equation}
\mathbb{L}^\dagger \mathbb{C}\mathbb{L}=\varpi\,,\label{Lsymp}
\end{equation}
where
\begin{equation}
\varpi \equiv -i\,\begin{pmatrix} {\bf 1} & {\bf 0} \cr {\bf 0} & -{\bf 1}\end{pmatrix}\,.
\end{equation}
If we change the complex index $\underline{N}$ into a real one by means of the Cayley matrix $\mathcal{A}$, thus defining:
\begin{equation}
\mathbb{L}_{{\rm Sp}}\equiv \mathbb{L}\mathcal{A}\,\,,\,\,\,\,\mathcal{A}\equiv \frac{1}{\sqrt{2}}\begin{pmatrix} {\bf 1} & i\,{\bf 1} \cr {\bf 1} & -i\,{\bf 1}\end{pmatrix}\,,
\end{equation}
Eq. (\ref{Lsymp}) expresses the condition that the real matrix $\mathbb{L}_{{\rm Sp}}$  be symplectic since $\varpi=\mathcal{A}\mathbb{C}\mathcal{A}^\dagger$. As a consequence of this also $\mathbb{L}_{{\rm Sp}}^T$ is symplectic and this implies an other set of identities which can be cast in the following compact form:
\begin{equation}
\mathbb{L}\varpi \mathbb{L}^\dagger=\mathbb{C}\,.
\end{equation}
In terms of $\mathbb{L}$ we define the following symmetric, negative-definite, symplectic matrix which encodes all information about the coupling of the vector fields to the scalars:
\begin{eqnarray}
\mathcal{M}(z,\bar{z})&=&(\mathcal{M}_{MN})\equiv \mathbb{C}\mathbb{L}\mathbb{L}^\dagger\mathbb{C}=\mathcal{M}(z,\bar{z})^T\,,\nonumber\\
\mathcal{M}\mathbb{C}\mathcal{M}&=&\mathbb{C}\,.
\end{eqnarray}
Under an isometry transformation $g:\,z\rightarrow z'$ in $G_{SK}$, using (\ref{OmK}), we find that $\mathcal{M}$ transforms linearly:
\begin{equation}
\mathcal{M}(z,\bar{z})\,\rightarrow\,\,\,\mathcal{M}(z',\bar{z}')=\mathbb{M}[g]^T\mathcal{M}(z,\bar{z})\mathbb{M}[g]\,.
\end{equation}
From the above properties of $V$ and $U_i$ we find the following general symplectic covariant relation:
\begin{equation}
U^{MN}\equiv g^{i\bar{\jmath}}\,U_i^M U_{\bar{\jmath}}^N=-\frac{1}{2}\mathcal{M}^{MN}-\frac{i}{2}\,\mathbb{C}^{MN}-\overline{V}^MV^N\,,\label{UMN}
\end{equation}
where $\mathcal{M}^{MN}$ are the components of $\mathcal{M}^{-1}=-\mathbb{L}\mathbb{L}^\dagger$.\par

If $k_a$ is the Killing vector defining an infinitesimal isometry, invariance of the K\"ahler form $K$, $\ell_a K=0$, implies:
\begin{equation}
\ell_aK=d(\iota_a K)=0\,\,\Rightarrow\,\,\,\,\iota_a K=-d \mathcal{P}_a\,,
\end{equation}
where $\iota_a$ denotes the contraction of $K$ with $k_a$. The last equation defines the momentum maps and is equivalent to Eqs. (\ref{kpal}).\par
The Killing vectors satisfy the Poisson-bracket relation:
\begin{equation}\label{lhs}
K\left(k_a,k_b\right)=ig_{i\bar{\jmath}}\,k^i_{[a}\,k^{\bar{\jmath}}_{b]}=ig^{k\bar{\jmath}}\partial_{\bar{\jmath}}\mathcal{P}_{[a}
\partial_{k}\mathcal{P}_{b]}\equiv  \frac{1}{2}\{\mathcal{P}_{a},\mathcal{P}_{b}\}= -\frac{1}{2}f_{ab}^c \mathcal{P}_{c}
\end{equation}
where the last equality was proven in \cite{D'Auria:1990fj}.

Finally let us prove equation (\ref{totti}).
To this aim, let us invert the metric in one of eq.s (\ref{kpal}):
\begin{equation}
g_{i\bar{\jmath}}\,k_a^i=i\,\partial_{\bar{\jmath}}\mathcal{P}_a\,,\label{kpalA}
\end{equation}
and use (\ref{kmetric}). Recalling the general condition on K\"ahler-manifold isometries $\partial_{\bar\jmath}\,k^i_a(z)=0$, we find:
\begin{equation}
\partial_{\bar\jmath}(k_a^i\,\partial_i\mathcal{K})=i\,\partial_{\bar{\jmath}}\mathcal{P}_a\,,
\end{equation}
which implies
\begin{equation}
k_a^i\, \partial_i\mathcal{K}=i\,\mathcal{P}_a\,+ C(z)\,. \label{pkA1}
\end{equation}
This would reproduce (\ref{totti}) if $C(z)=f(z)$. To fix the holomorphic function $C(z)$, it is sufficient to consider the holomorphic derivative of (\ref{Lie1}), which implies:
\begin{equation}
g_{j\bar\jmath}k^{\bar\jmath}_a+\partial_j(k^i_a\partial_i \mathcal{K})= -\partial_j f_a\,,
\end{equation}
that is, using (\ref{kpal}):
\begin{equation}
-i \,\partial_j \mathcal{P}_a\,+\partial_j(k^i_a\partial_i \mathcal{K})= -\partial_j f_a\,. \label{pfA}
\end{equation}
By inserting now (\ref{pkA1}) in (\ref{pfA}), one finally finds the  identification $C(z)=f(z)$, modulo an additive constant that, as discussed in section 2, can be absorbed in the definition of $\mathcal{P}_a$.

%
%
%

\paragraph{Quaternionic K\"ahler Manifolds} Here we briefly recall the definition of a quaternionic  K\"ahler manifold\footnote{We shall be interested in non-compact quaternionic K\"ahler manifolds with negative curvature as only these are relevant to supergravity.} $\mathcal{M}_{QK}$ \cite{Bagger:1983tt,Hitchin:1986ea,Galicki:1986ja,D'Auria:2001kv}  and fix the notations.
$\mathcal{M}_{QK}$ is a $4n_H$-dimensional real, Riemannian manifold with holonomy group:
\begin{equation}
H={\rm SU}(2)\times H'\,\,,\,\,\,\,H'\subset {\rm Sp}(2n_H,\mathbb{R})\,,\label{Hsu2Hp}
\end{equation}
where ${\rm SU}(2)$, together with the group ${\rm U}(1)$ of  K\"ahler transformations in the holonomy group of $\mathcal{M}_{SK}$, define the ${\rm U}(2)$ R-symmetry group of the supersymmetry algebra.\par
 The positive definite metric is denoted by $h_{uv}(q)$, where $q^u$ are the coordinates describing the scalar fields of the hypermultiplets.
The action of the ${\rm SU}(2)$ generators on the tangent space defines three complex structures $J^x{}^u{}_v$, $x=1,2,3$, satisfying the quaternionic algebra:
\begin{equation}
J^xJ^y=-\delta^{xy}+\epsilon^{xyz}\,J^z\,.\label{Jstruc}
\end{equation}
In terms of this quaternionic structure, a triplet of hyper-K\"ahler 2-forms are defined:
\begin{equation}
K^x=K^x_{uv}\,dq^u\wedge dq^v\,\,,\,\,\,K^x_{uv}=h_{uw}\,J^x{}^w{}_v\,.
\end{equation}
The above definition and Eq. (\ref{Jstruc}) imply the following relation:
\begin{equation}
K^x_{uw}h^{ws}K_{sv}^y=-\delta^{xy}\,h_{uv}+\epsilon^{xyz}\,K^z_{uv}\,,\label{Kstruc}
\end{equation}
where, as usual, $h^{uv}$ are the components of the inverse metric. One of the defining properties of quaternionic K\"ahler manifolds is that
$K^x$ be \emph{covariantly constant} with respect to the ${\rm SU}(2)$-connection $\omega^x$:
\begin{equation}
\nabla K^x=dK^x+\epsilon^{xyz}\,\omega^y\wedge  K^z=0\,.\label{DK0}
\end{equation}
In terms of the connection 1-forms $\omega^x$ we define the ${\rm SU}(2)$-curvature $\Omega^x$:
\begin{equation}
\Omega^x\equiv d\omega^x+\frac{1}{2}\epsilon^{xyz}\,\omega^y\wedge \omega^z\,,
\end{equation}
The other defining property of a quaternionic K\"ahler manifold is that the hyper-K\"ahler 2-forms be proportional to the ${\rm SU}(2)$-curvature:
\begin{equation}
\Omega^x=\lambda \,K^x\,,\label{OmKx}
\end{equation}
where $\lambda$ is a real coefficient depending on the normalization of the metric.
Choosing the standard normalization of the kinetic term
for the hyperscalars $q^u$ amounts to fixing $\lambda=-1$. The above equation is consistent with (\ref{DK0}) by virtue of the covariant constancy of  $\Omega^x$:
\begin{equation}
\nabla \Omega^x=d\Omega^x+\epsilon^{xyz}\,\omega^y\wedge \Omega^z=0\,.
\end{equation}
Property (\ref{Hsu2Hp}) implies that we can define the vielbein 1-forms as follows:
\begin{equation}
\mathcal{U}^{A\alpha}=\mathcal{U}^{A\alpha}_u\,dq^u\,,
\end{equation}
where $A=1,2$ is the ${\rm SU}(2)$-doublet index labeling the supersymmetries and $\alpha=1,\dots ,2 n_H$ labels the fundamental representation of
${\rm Sp}(2n_H,\mathbb{R})$. In this basis the rigid tangent space index ${\bf u}$ is a composite one ${\bf u}=(A,\alpha)$ and the rigid metric is $\eta_{{\bf u}{\bf v}}=\epsilon_{AB}\mathbb{C}_{\alpha\beta}$, where $\mathbb{C}_{\alpha\beta}$ is the ${\rm Sp}(2n_H,\mathbb{R})$-invariant matrix, so that:
\begin{equation}
\mathcal{U}^{A\alpha}_u\mathcal{U}^{B\beta}_u\epsilon_{AB}\,\mathbb{C}_{\alpha\beta}=h_{uv}\,.
\end{equation}
These 1-forms satisfy the following relations which we shall need in our discussion:
\begin{eqnarray}
\mathcal{U}_{A\alpha}&\equiv &(\mathcal{U}^{A\alpha})^*=\epsilon_{AB}\mathbb{C}_{\alpha\beta}\,\mathcal{U}^{B\beta}\,,\nonumber\\
\mathcal{U}_{A\alpha\,u}\,\mathcal{U}^{B\alpha}_{v}&=&\frac{1}{2}\,h_{uv}\,\delta_A^B-\frac{i}{2}\,K^x_{uv}\,(\sigma^x)_A{}^B\,,\label{UUKs}
\end{eqnarray}
where the relative sign between the two terms on the right hand side of last equation is fixed by (\ref{Kstruc}).
Moreover the vielbein 1-forms are covariantly constant, namely the satisfy the condition:
\begin{equation}
\nabla \mathcal{U}^{A\alpha}\equiv d\mathcal{U}^{A\alpha}+\frac{i}{2}\,(\sigma^x)_B{}^A\,\omega^x\wedge \mathcal{U}^{B\alpha}+\Delta^{\alpha\gamma}\wedge \mathcal{U}^{A\beta}\mathbb{C}_{\gamma\beta}=0\,,
\end{equation}
where $\Delta^{\alpha\beta}=\Delta^{\beta\alpha}$ denote the $H'\subset Sp(2n_H,\mathbb{R})$-connection 1-forms.\par
The Riemann tensor of a quaternionic manifold has the general form:
\begin{equation}
\mathcal{R}_{uv|ts}=\frac{i}{2}\, (\sigma^x)_{A}{}^B\,\Omega^x_{ts}\,\mathcal{U}_u^{A\alpha}\mathcal{U}_{B\alpha|v}+ \mathbb{R}_{\alpha\beta|ts}\,\mathcal{U}_u^{A\alpha}\mathcal{U}^\beta_{A|v}\,.
\end{equation}
$\mathbb{R}_{\alpha\beta}$ denotes instead the $H'\subset Sp(2n_H,\mathbb{R})$-curvature, defined in terms of the connection one-form $\Delta^{\alpha\beta}$ as follows
\begin{equation}
\mathbb{R}_{\alpha\beta}\equiv d\Delta^{\alpha\beta}+\mathbb{C}_{\gamma\delta}\Delta^{\alpha\gamma}\wedge\Delta^{\delta\beta}\,.
\end{equation}

Consider now infinitesimal isometries generated by $t_m$, whose action on the scalar fields is described by Killing vectors $k_m=k_m^u\,\partial_u$. They close the isometry algebra:
\begin{equation}
[t_m,\,t_n]=f_{mn}{}^p\,t_p\,\,\,,\,\,\,\,\,[k_m,\,k_n]=-f_{mn}{}^p\,k_p\,,
\end{equation}
and leave the 4-form $\sum_{x=1}^3 K^x\wedge K^x$ invariant \cite{D'Auria:1990fj}. This condition amounts to requiring:
\begin{equation}
\ell_n K^x=\epsilon^{xyz}\,K^y\,W^z_n\,,\label{elKx}
\end{equation}
where $W^z_n$ is an ${\rm SU}(2)$-compensator. Equation (\ref{elKx}) is solved by writing the Killing vectors $k_n$ in terms of \emph{tri-holomorphic momentum maps} $\mathcal{P}_n^x$ as follows \cite{D'Auria:1990fj}:
\begin{equation}
\iota_n K^x=-\nabla \mathcal{P}^x_n=-(d\mathcal{P}^x_n+\epsilon^{xyz}\omega^y\,\mathcal{P}_n^z)\,,\label{inKx}
\end{equation}
provided
\begin{equation}
 \mathcal{P}^x_n=\lambda^{-1}(\iota_n \omega^x-W^x_n)=W^x_n-\iota_n \omega^x\,,\label{PHW}
\end{equation}
where we have used $\lambda=-1$. The above equation was derived in \cite{Galicki:1986ja}, see also \cite{D'Auria:1990fj}.
For those isometries with vanishing compensator, $W^x_n=0$, the momentum maps have the simple expression: $\mathcal{P}^x_n=-k^u_n\,\omega^x_u$.\par
Just has for the special K\"ahler manifolds, (see equation (\ref{lhs})), the momentum maps satisfy Poisson brackets described by the following equivariance condition:
\begin{equation}
2\,K_{uv}\,k^u_n\,k^v_m-\lambda\,\epsilon^{xyz}\,\mathcal{P}_n^y\,\mathcal{P}_m^z=-f_{mn}{}^p\,\mathcal{P}_p^x\,.\label{equivar2}
\end{equation}
For homogeneous symmetric manifolds $k_n$ and $\mathcal{P}^x_n$ can be given a simple geometric characterization.
Indeed if $\mathcal{M}_{QK}$ has the general form:
\begin{equation}
\mathcal{M}_{QK}=\frac{G_{qk}}{H}\,,
\end{equation}
where  $G_{qk}$ is the isometry group, denoting by $\mathfrak{g}_{qk}$ and $\mathfrak{H}$ the Lie algebras of $G_{qk}$ and  $H$, respectively,
we can write the Cartan decomposition of  $\mathfrak{g}_{qk}$ into compact and non-compact generators:
\begin{equation}
\mathfrak{g}_{qk}=\mathfrak{H}\oplus\mathfrak{K}\,,
\end{equation}
where $[\mathfrak{H},\,\mathfrak{H}]\subset \mathfrak{H}$, $[\mathfrak{H},\,\mathfrak{K}]\subset \mathfrak{K}$ and $[\mathfrak{K},\,\mathfrak{K}]\subset \mathfrak{H}$ (symmetry).
The coset space $\mathfrak{K}$   is generated by a basis of non-compact generators $K_{{\bf u}}$, ${\bf u}=1,\dots, 4n_H$ be the rigid tangent space index.
The generators of $H$ split into the generators $J^x$ of ${\rm SU}(2)$ and $J_{\alpha\beta}=J_{\beta\alpha}$ of $H'$, according to the decomposition (\ref{Hsu2Hp}).
The symmetry property of the manifold implies $[\mathfrak{K},\,\mathfrak{K}]\subset \mathfrak{H}$, or, in components:
\begin{equation}
[K_{{\bf u}},\,K_{{\bf v}}]=f_{{\bf u}{\bf v}}{}^{x}\,J^x+\frac 12 f_{{\bf u}{\bf v}}{}^{\alpha\beta}\,J_{\alpha\beta}\,.\label{KKJ}
\end{equation}
We can normalize the generators so that the Cartan-Killing form $(\,,\,)$ of $\mathfrak{g}_{qk}$ is
\begin{equation}
(K_{{\bf u}},\,K_{{\bf v}})=\delta_{{\bf u}{\bf v}}\,\,,\,\,\,\,(J^x,\,J^y)=-\delta^{xy}\,, \,\,\,\,(J_{\alpha\beta},\,J_{\gamma\delta})=-2 \,\mathbb{C}_{\alpha(\gamma}\mathbb{C}_{\delta)\beta}\,.
\end{equation}
The vielbein and connections are, as usual, defined by decomposing the left invariant one-form in components along $\mathfrak{K}$ and
 $\mathfrak{H}$:
 \begin{equation}
\Gamma= L^{-1}dL=V^{{\bf u}}\,K_{{\bf u}}+\frac{1}{2}\omega^x\,J^x+\frac 12 \Delta^{\alpha\beta}\,J_{\alpha\beta}\,,\label{li1f}
 \end{equation}
where $L$ is the coset representative in some representation of $G_{qk}$, so that
\begin{equation}
V^{{\bf u}}=(K_{{\bf u}},\,\Gamma)\,\,,\,\,\,\,\omega^x=-2\,(J^x,\,\Gamma)\,,\,\,\,\,\Delta^{\alpha\beta}= (J^{\alpha\beta},\,\Gamma)\,.
\end{equation}
From the Maurer-Cartan equations $d\Gamma+\Gamma\wedge \Gamma=0$ we can read off the expression for the curvature and the 2-forms $K^x$:
\begin{equation}
\Omega^x=d\omega^x+\frac{1}{2}\epsilon^{xyz}\,\omega^y\wedge \omega^z=-f_{{\bf u}{\bf v}}{}^{x}V^{{\bf u}}\wedge V^{{\bf v}}=-K^x\,,
\end{equation}
where we have used (\ref{KKJ}) and (\ref{OmKx}) with $\lambda=-1$. From this we derive the holonomic components of $K^x$:
\begin{equation}
K^x_{uv}=f_{{\bf u}{\bf v}}{}^{x}V_u{}^{{\bf u}}\, V_v{}^{{\bf v}}\,.\label{Kxuvsym}
\end{equation}
We can give the following useful characterization of the Killing vector $k_n$ and the momentum map $\mathcal{P}_n^x$ associated with the isometry generator $t_n\in \mathfrak{g}_{qk}$:
\begin{equation}
L^{-1}t_n L=k_n^u\,V_{u}{}^{{\bf u}}\,K_{{\bf u}}-\frac{1}{2}\mathcal{P}_n^x\,J^x+\frac 12 \Sigma_n^{\alpha\beta}\,J_{\alpha\beta}\,.\label{Lm1tL}
\end{equation}
We prove below that $k_n$ and  $\mathcal{P}_n^x$  defined in (\ref{Lm1tL}) do satisfy (\ref{inKx}).
From (\ref{Kxuvsym}) and (\ref{li1f}) we find:
\begin{equation}
2 k_n^u\,K_{uv}^x=2\,f_{{\bf u}{\bf v}}{}^{x}\,k_n^u\,V_u{}^{{\bf u}}\, V_v{}^{{\bf v}}=-2\,([L^{-1} t_n L,\,L^{-1}\partial_v L],J^x)+\epsilon^{xyz}\,\mathcal{P}^y_n \omega^z_v\,.\label{equa1q}
\end{equation}
Now let us evaluate $\nabla \mathcal{P}_n^x$:
\begin{eqnarray}
\nabla_v \mathcal{P}_n^x&=&\partial_v \mathcal{P}_n^x+\epsilon^{xyz}\omega_v^y\,\mathcal{P}_n^z=2\,(\partial_vL^{-1}t_nL+L^{-1}t_n \partial_v L,J^x)+\epsilon^{xyz}\omega_v^y\,\mathcal{P}_n^z=\nonumber\\
&=&2\,([L^{-1} t_n L,\,L^{-1}\partial_v L],J^x)+\epsilon^{xyz}\omega_v^y\,\mathcal{P}_n^z=-2 k_n^u\,K_{uv}^x\,,
\end{eqnarray}
where in the last equality we have used (\ref{equa1q}).

 Let us now prove (\ref{PHW}). From basic coset geometry we know that the left action of an isometry on the coset representative $L$ yields $L$ computed in the transformed point, multiplied to the right by a compensator in $H$. For an infinitesimal isometry this
is expressed by the property:
\begin{equation}
t_n\,L=k_n^u\,\partial_uL+L\,W_n\,.\label{tnL}
\end{equation}
where $W_n\in \mathfrak{H}$ is the infinitesimal generator of the compensating transformation, which can be expanded as follows
\begin{equation}
W_n=-\frac{1}{2}\,W_n^xJ^x+\frac 12 W_n^{\alpha\beta}\,J_{\alpha\beta}\,.
\end{equation}
Multiplying (\ref{tnL}) to the left by $L^{-1}$ we find:
\begin{equation}
L^{-1}t_nL=k_n^u\,\Gamma_u+W_n=k_n^u\,V_u{}^{{\bf u}}\,K_{{\bf u}}+\frac{1}{2}k_n^u\,\omega_u^x\,J^x+\frac 12 k_n^u\,\omega_u^{\alpha\beta}\,J_{\alpha\beta}
-\frac{1}{2}\,W_n^xJ^x+\frac 12 W_n^{\alpha\beta}\,J_{\alpha\beta}\,.
\end{equation}
Comparing the above expansion with (\ref{Lm1tL}) we find:
\begin{equation}
\mathcal{P}_n^x=W_n^x-k_n^u\,\omega_u^x\,,
\end{equation}
which is (\ref{PHW}). Equations (\ref{PHW}) and (\ref{inKx}) then imply (\ref{elKx}).\par
Consider now a \emph{solvable} (or Iwasawa) parametrization of the coset for which we describe the quaternionic K\"ahler manifold as globally isometric
to a solvable Lie group generated by a solvable Lie algebra $Solv$ \cite{solvable}:
\begin{equation}
\mathcal{M}_{QK}\sim \exp{(Solv)}\,.
\end{equation}
The coset representative is then an element of $\exp{(Solv)}$:
\begin{equation}
L(q)=e^{q^u\,T_u}\in \exp{(Solv)}\,,
\end{equation}
where $T_u$ are the generators of $Solv$. Being $L(q)$ an element of a group, the action on it of any other element of the same group has no compensating transformation:
\begin{equation}
\forall g\in \exp{(Solv)}\,\,:\,\,\,\,\,gL(q)=L(q')\,.
\end{equation}
Therefore for any $t_n\in Solv$ we have $W_n=0$, i.e.
\begin{equation}
\mathcal{P}_n^x=-k_n^u\,\omega^x_u\,.
\end{equation}
Transformations in $\exp(Solv)$ comprise \emph{translational isometries}.

\section{Proofs of some symplectically-covariant relations on the gauging}
\label{ident}

Let us prove here  the  identities (\ref{newidentities}):
\begin{equation}
\mathcal{P}_M\Omega^M=0\,\,,\,\,\,k_M^i\,\Omega^M=0\,.\label{newidentitiesapp}
\end{equation}
To prove the first one we write (\ref{PVtV}) for the gauge-momentum maps:
\begin{equation}
\mathcal{P}_M=-e^\mathcal{K}\,{\rm X}_{MNP}\overline{\Omega}^N\Omega^P\,.
\end{equation}
Contracting both sides with $\Omega^M$ we find:
\begin{equation}
\Omega^M\mathcal{P}_M=-e^\mathcal{K}\,\Omega^M {\rm X}_{MNP}\overline{\Omega}^N\Omega^P=\frac{e^\mathcal{K}}{2}\, \overline{\Omega}^N{\rm X}_{NMP}\Omega^M\Omega^P=0\,,\label{OmP}
\end{equation}
where we have used the linear constraint (\ref{lc}) and the symplectic property of the matrices ${\rm X}_{MN}{}^P$:
\begin{equation}
2{\rm X}_{(MP)N}=-{\rm X}_{NMP}\,,
\end{equation}
being ${\rm X}_{MNP}\equiv {\rm X}_{MN}{}^Q\mathbb{C}_{QP}$. Last equality in (\ref{OmP}) then follows from (\ref{tOmOm}).\par
 Let us now prove the second of (\ref{newidentitiesapp})
\begin{equation}
\Omega^M\,k_M^i=i\,g^{i\bar{\jmath}}\,\Omega^M\,\partial_{\bar{\jmath}} \mathcal{P}_M=i\,g^{i\bar{\jmath}}\,\partial_{\bar{\jmath}}(\Omega^M\, \mathcal{P}_M)=0\,,
\end{equation}
where we have used the first of (\ref{newidentitiesapp}).

From (\ref{newidentitiesapp}) we can deduce the following relations:
\begin{equation}
D_i(V^M \mathcal{P}_M)=0\,\,\Rightarrow\,\,\,U_i^M \mathcal{P}_M+V^M\partial_i\mathcal{P}=0\,\,\Rightarrow\,\,\,
U_i^M \mathcal{P}_M+i\,g_{i\bar{\jmath}}\,k_M^{\bar{\jmath}}V^M=0\,.
\end{equation}
Contracting (\ref{kaUi}) with the embedding tensor we find:
\begin{equation}
k_M^i\,U_i^P=-{\rm X}_{MN}{}^P\,V^N+i\,\mathcal{P}_M\,V^P\,.\label{kMUi}
\end{equation}
Contracting both sides with $\overline{V}^M$ and using the first of (\ref{newidentitiesapp}) we find:
\begin{equation}
\overline{V}^Mk_M^i\,U_i^P=-{\rm X}_{MN}{}^P\,\overline{V}^M V^N\,.
\end{equation}
Next we contract both sides with $\Theta_P$, where $\Theta_P$ can be either $\Theta_P{}^a$ or $\Theta_P{}^n$ and use the quadratic constraints
(\ref{XXXX}) which imply that the \emph{generalized structure constants} ${\rm X}_{MN}{}^P$ are antisymmetric in the first two indices only if contracted
to the right by $\Theta_P$: ${\rm X}_{MN}{}^P\Theta_P=-{\rm X}_{NM}{}^P\Theta_P$. By virtue of this feature we find:
\begin{equation}
\overline{V}^Mk_M^i\,U_i^P\Theta_P=-{\rm X}_{MN}{}^P\,\overline{V}^M V^N\Theta_P={\rm X}_{NM}{}^P\,\overline{V}^M V^N\Theta_P=-{V}^Mk_M^{\bar{\imath}}\,\overline{U}_{\bar{\imath}}^P\Theta_P\,.\label{kMUi2app}
\end{equation}

\paragraph{The general Ward identity}

Let us now prove the Ward identity \cite{ward} for the generic dyonic gauging of $\mathcal{N}=2$ supergravity.
We shall evaluate each term in the left hand side of (\ref{Ward}) separately. From the above definitions we find:
\begin{eqnarray}
W^{i\,AC} \overline{W}^{\bar{\jmath}}_{BC}g_{i\bar{\jmath}}&=&\delta_B^A\,k_M^ik_N^{\bar{\jmath}}g_{i\bar{\jmath}}\overline{V}^M\,V^N-i\,(\sigma^x)_B{}^A\,\left(
k_M^{\bar{\jmath}}\,V^M\,\overline{U}_{\bar{\jmath}}^N-k_M^{i}\,\overline{V}^M\,{U}_{i}^N\right)\,\mathcal{P}^x_N+\nonumber\\
&&+(\sigma^x\sigma^y)_B{}^A\,\mathcal{P}^x_M\mathcal{P}^y_N U^{MN}\,,
\end{eqnarray}
where $U^{MN}\equiv {U}_{i}^N\,g^{i\bar{\jmath}}\,\overline{U}_{\bar{\jmath}}^N$, see (\ref{UMN}). On the right hand side of the above expression we split the terms proportional to $\delta_B^A$ from those proportional to $(\sigma^x)_B{}^A$ and use Eq. (\ref{kMUi2}) to find:
\begin{eqnarray}
W^{i\,AC} \overline{W}^{\bar{\jmath}}_{BC}g_{i\bar{\jmath}}&=&\delta_B^A\,\left(k_M^ik_N^{\bar{\jmath}}g_{i\bar{\jmath}}\overline{V}^M\,V^N+\mathcal{P}^x_N\mathcal{P}^x_M U^{MN}\right)+i\,(\sigma^x)_B{}^A\,\left(-2\,
{\rm X}_{MN}{}^P\overline{V}^M\,V^N\,\mathcal{P}^x_P+\right.\nonumber\\&&+\left.\epsilon^{xyz}\,\mathcal{P}^y_M\mathcal{P}^z_N U^{[MN]}\right)\,.
\end{eqnarray}
Now use Eqs. (\ref{UMN}) and the locality constraint (\ref{qc3}) to write:
\begin{equation}
\mathcal{P}^y_M\mathcal{P}^z_N U^{[MN]}=-\frac{i}{2}\,\mathcal{P}^y_M\mathcal{P}^z_N \mathbb{C}^{MN}-\mathcal{P}^y_M\mathcal{P}^z_N \overline{V}^{[M} V^{N]}=-\mathcal{P}^y_M\mathcal{P}^z_N \overline{V}^{[M} V^{N]}\,,
\end{equation}
so that we finally find:
\begin{eqnarray}
W^{i\,AC} \overline{W}^{\bar{\jmath}}_{BC}g_{i\bar{\jmath}}&=&\delta_B^A\,\left(k_M^ik_N^{\bar{\jmath}}g_{i\bar{\jmath}}\overline{V}^M\,V^N+\mathcal{P}^x_N\mathcal{P}^x_M U^{MN}\right)+i\,(\sigma^x)_B{}^A\,\left(-2\,
{\rm X}_{MN}{}^P\overline{V}^M\,V^N\,\mathcal{P}^x_P+\right.\nonumber\\&&-\left.\epsilon^{xyz}\,\mathcal{P}^y_M\mathcal{P}^z_N \, \overline{V}^{M} V^{N}\right)
\end{eqnarray}
Let us now move to the evaluation of the square of the hyperini shifts:
\begin{eqnarray}
2\,N_\alpha{}^A\,\overline N^\alpha{}_A=8\,\mathcal{U}^{A\alpha}_u\,\mathcal{U}_{v\,B\alpha}\,k^u_M\,k^v_N\,\overline{V}^M{V}^N=4\left(\delta_B^A h_{uv}+i\,(\sigma^x)_B{}^A\,K^x_{uv}\right)k^u_M\,k^v_N\,\overline{V}^M{V}^N\,.\label{NNesp}
\end{eqnarray}
where we have used Eq. (\ref{UUKs}). Finally let us compute the square of the gravitini shifts:
\begin{align}
-12\,\overline S^{AC}\,S_{BC}&=-3\,(\sigma^x\sigma^y)_B{}^A\,\mathcal{P}^x_M\mathcal{P}^y_N \,V^M\overline{V}^N=-3\,\mathcal{P}^x_M\mathcal{P}^x_N\,V^M\overline{V}^N+3i\,\epsilon^{xyz}\,\mathcal{P}^y_M\mathcal{P}^z_N \, \overline{V}^{M} V^{N}(\sigma^x)_B{}^A\,.\nonumber\\
&\label{SSesp}
\end{align}
We can now compute the left hand side of the Ward identity:
\begin{equation}
g_{i\bar{\jmath}}\,W^{i\,AC}\overline{W}_{BC}^{\bar{\jmath}}+2\,N_\alpha{}^A\,\overline N^{\alpha}{}_{B}-12 \,\overline S^{AC}S_{BC}=\delta_B^A\,V(z,\bar{z},q)+i\,Z^x\,(\sigma^x)_B{}^A\,,
\end{equation}
where
\begin{equation}
V(z,\bar{z},q)=(k_M^ik_N^{\bar{\jmath}}g_{i\bar{\jmath}}+4\,h_{uv}k_M^uk_N^v)\overline{V}^M\,V^N+(U^{MN}-3\,V^M\overline{V}^N)\mathcal{P}^x_N\mathcal{P}^x_M \,, \label{potentialVapp}
\end{equation}
is the general symplectic invariant expression of the scalar potential given in \cite{deWit:2005ub} as a generalization to dyonic gaugings of the one given in \cite{Andrianopoli:1996cm}, and
\begin{equation}
Z^x=(-2\,
{\rm X}_{MN}{}^P\,\mathcal{P}^x_P+2\,\epsilon^{xyz}\,\mathcal{P}^y_M\mathcal{P}^z_N+4\,K^x_{uv}k^u_M\,k^v_N)\overline{V}^{M} V^{N}\,.
\end{equation}
From the equivariance condition (\ref{gequiv2}) it follow that $Z^x=0$, so that the Ward identity is proven.

\section{Rescalings}

Let us summarize here the relation between the couplings and fields of the rigid-supersymmetric thery, identified with an upper ring, and the corresponding supergravity fields. We find that the  resscaling only affects the vector-multiplet sector, and in particular the gaugini: \begin{equation}\label{gau}
   \mathring \lambda^{iA}=\frac 1 \mu  \lambda^{iA}
\end{equation}
 the special geometry sector, in a generic coordinate frame:
\begin{eqnarray}\label{vring2}
  V^{M}&=& \left(
\begin{array}
[c]{c}%
X^0 \\
0 \\
F_0   \\
0
\end{array}
\right)\,+ \frac 1\mu \left(
\begin{array}
[c]{c}%
0 \\
 \mathring{X}^I (z,\bar z) \\
0  \\
\mathring{F}_I (z,\bar z)
\end{array}
\right) \,+O\left(  1/\mu
^{2}\right) \,;
\\\label{uringapp}
\tilde U^{M}_i&=&\frac 1\mu\left(
\begin{array}
[c]{c}%
0 \\
\partial_i   \mathring{X}^I \equiv \mathring f^I_i \\
0   \\
\partial_i   \mathring{F}_I \equiv \mathring h_{Ii}
\end{array}
\right) \,+O\left(  1/\mu
^{2}\right) \,,
\end{eqnarray}
from which we get, in the limit $\mu \to \infty$:
\begin{align*}
g_{i\bar{\jmath}}  & \to \frac{1}{\mu^{2}}\mathring{g}_{i\bar{\jmath}%
}\text{ \ \ \ \ \ \ \ \ \ \ }C_{ijk}\to \frac{1}{\mu^{2}}\mathring C_{ijk}\\
R_{i\bar{\jmath}k\bar{l}}  & \to \frac{1}{\mu^{2}}\mathring R_{i\bar{\jmath
}k\bar{l}} \\
\Gamma_{jk}^{i}  &  \rightarrow\mathring \Gamma_{jk}^{i}\text{
\ \ \ \ \ \ \ \ \ \ \ \ \ }\mathcal{\ Q\rightarrow}\frac{1}{\mu^{2}%
}\mathring{\mathcal{Q}}\,,
\end{align*}
together with the embedding tensor:
\begin{equation}\label{em}
   \Theta^{m}_M = \frac 1{\mu^2}\mathring \Theta^{m}_M\,.
\end{equation}

\end{document}